\documentclass[twocolumn]{aastex62}
\usepackage{xspace}
\usepackage{amsmath}
\newcommand{\nraogboblurb}{The Green Bank Observatory and the National Radio Astronomy Observatory are
facilities of the National Science Foundation operated under cooperative
agreement by Associated Universities, Inc.}
\newcommand{\hide}[1]{}

\newcommand{\lsim}{\ensuremath{\,\lesssim\,}}
\newcommand{\gl}{\ensuremath{\ell}}
\newcommand{\gb}{\ensuremath{{\it b}}}
\newcommand{\absb}{\ensuremath{\vert\,\gb\,\vert}}

\newcommand{\lb}{\ensuremath{(\gl,\gb)}}

\newcommand{\sgra}{Sgr\thinspace A\ensuremath{^*}}
\newcommand{\sgramath}{\ensuremath{\rm SgrA^{*}}}
\newcommand{\kms}{\ensuremath{\,{\rm km\,s^{-1}}}}

\newcommand{\pc}{\ensuremath{\,{\rm pc}}}
\newcommand{\kpc}{\ensuremath{\,{\rm kpc}}}

\newcommand{\degree}{\ensuremath{\,^\circ}}

\newcommand{\rgal}{\ensuremath{\,R_{\rm Gal}}}   

\newcommand{\hi}{{\rm H\,{\footnotesize I}}}
\newcommand{\hii}{{\rm H\,{\footnotesize II}}}
\newcommand{\cii}{{\rm C\,{\footnotesize II}}}

\newcommand{\methanol}{\ensuremath{\rm CH_3OH}}

\newcommand{\thetat}{\ensuremath{\theta_{\rm tilt}}}
\newcommand{\thetar}{\ensuremath{\theta_{\rm roll}}}

\shorttitle{Galactic midplane with \hii\ Regions}
\shortauthors{Anderson et al.}
\begin{document}

\title{A Galactic Plane Defined by the Milky Way \hii\ Region Distribution}

\author{L.~D.~Anderson}
\affiliation{Department of Physics and Astronomy, West Virginia University, Morgantown WV 26506, USA}
\affiliation{Adjunct Astronomer at the Green Bank Observatory, P.O. Box 2, Green Bank WV 24944, USA}
\affiliation{Center for Gravitational Waves and Cosmology, West Virginia University, Chestnut Ridge Research Building, Morgantown, WV 26505, USA}

\author{Trey~V.~Wenger}
\affiliation{Astronomy Department, University of Virginia, P.O. Box 400325, Charlottesville, VA 22904-4325, USA}

\author{W.~P.~Armentrout}
\affiliation{Department of Physics and Astronomy, West Virginia University, Morgantown WV 26506, USA}
\affiliation{Center for Gravitational Waves and Cosmology, West Virginia University, Chestnut Ridge Research Building, Morgantown, WV 26505, USA}
\affiliation{Green Bank Observatory, P.O. Box 2, Green Bank WV 24944, USA}

\author{Dana~S.~Balser}
\affiliation{National Radio Astronomy Observatory, 520 Edgemont Road, Charlottesville VA, 22903-2475, USA}

\author{T.~M.~Bania}
\affiliation{Institute for Astrophysical Research, Department of Astronomy, Boston University, 725 Commonwealth Ave., Boston MA 02215, USA}

\begin{abstract}
We develop a framework for a new definition of the Galactic midplane, allowing for tilt (\thetat; rotation about Galactic azimuth 
$90\degree$), and roll (\thetar; rotation about Galactic azimuth
$0\degree$) of the midplane with respect to the current
definition.  Derivation of the tilt and roll angles also determines the solar height above the midplane.  Here we use nebulae from the {\it
  WISE} Catalog of Galactic \hii\ Regions
to
  define the Galactic high-mass star formation (HMSF) midplane. 
We analyze various subsamples of the {\it WISE} catalog and find that all have Galactic latitude scale heights near $0.30\degree$ and $z$-distribution scale heights near 30\,\pc.
The vertical distribution for small (presumably young) \hii\ regions is narrower than that of larger (presumably old) \hii\ regions ($\sim 25\,\pc$ versus $\sim 40\,\pc$), implying that the larger regions have migrated further from their birth sites. 
For all \hii\ region subsamples and for a variety of fitting methodologies, we find that the HMSF midplane is not significantly tilted or rolled with respect  to the currently-defined midplane, and therefore the Sun is near to the HMSF midplane.
These results are consistent with other studies of HMSF, but are inconsistent with many stellar studies, perhaps due to asymmetries in the stellar distribution near the Sun.  
Our results are sensitive to latitude restrictions, and also to the
completeness of the sample, indicating that similar analyses cannot be
done accurately with less complete samples.  The midplane framework we develop can be used for any future sample of Galactic objects to redefine the midplane.  

\end{abstract}

\keywords{Galaxy: structure -- ISM: \hii\ regions}

\section{Introduction\label{sec:intro}}

The midplane, the plane at Galactic latitude $b = 0\degree$, was defined in 1958
by the IAU subcommission 33b, which set the Galactic coordinate system
\citep{blaauw60}.  The IAU
  midplane definition comes from the Galactic Center location in B1950
  coordinates of (17:42:26.6, $-$28:55:00) and the north Galactic pole
  location in B1950 coordinates of (12:49:00, +27:24:00).  Ideally, the midplane definition would contain the 
minimum of the Galactic potential and there would be equal amounts of material above and below the midplane. The vertical distribution of objects with respect to the Galactic
midplane tells us fundamental parameters of Galactic structure, such
as the scale height of the objects studied, the Sun's height above or
below the midplane, $z_\odot$,  and even the orientation of the midplane itself.
Nearly all previous studies of the vertical
distribution of objects in the Galaxy have found an asymmetry in the
distribution of sources above and below the plane, with more sources
found below the IAU plane than above it.  This asymmetry is generally assumed to be the result of the Sun's location above the IAU Galactic midplane.

\begin{table*}
  \centering
  \setlength{\tabcolsep}{2pt}
  \begin{scriptsize}
  \begin{centering}
    \caption{Previous Results Analyzing the $b$ and $z$-Distributions of Galactic Objects\label{tab:previous}}
    \begin{tabular}{lccccccl}
      \hline
Tracer					& \multicolumn{2}{c}{Galactic Latitude $b$} && \multicolumn{2}{c}{Height $z$}&&\\ \cline{2-3} \cline{5-6}
&Scale Height$^{\rm a}$     &Peak&&Scale Height$^{\rm a}$     &Peak& Zone &Reference$^{\rm b}$  \\
					&(deg.)    &(deg.)    && (pc) 	      	    & (pc)	    		&&\\ \hline
ATLASGAL 870\,\micron\ cont. 	& $0.3$ 		& $-0.076\pm0.008$ 	&& && $-60\degree < \ell < 60\degree; \absb \le 1.5\degree$ & ~~~~~~1\\ 
ATLASGAL 870\,\micron\ cont.	& 			&&& $28\pm2$	    &$-6.7\pm1.1$    	& $-60\degree < \ell < 60\degree; \absb \le 1.5\degree$ & ~~~~~~2 \\
ATLASGAL 870\,\micron\ cont.	& 			&&& $31\pm3$	    &$-4.1\pm1.7$	& $-60\degree < \ell < 0\degree; \absb \le 1.5\degree$ & ~~~~~~2 \\
ATLASGAL 870\,\micron\ cont.	& 			&&& $24\pm1$	    &$-10.3\pm0.5$	& $0\degree < \ell < 60\degree; \absb \le 1.5\degree$\phn & ~~~~~~2\\
BGPS 1.1\,mm cont.		& 			& $-0.095\pm0.001$ 	&& 		    && $-10\degree < \ell < 90.5\degree; \absb < 0.5\degree$ & ~~~~~~3 \\
BGPS 1.1\,mm cont.		&			&&& $27\pm1$	    &$-9.7\pm0.6$	& $15\degree < \ell < 75\degree; \absb < 0.5\degree$							& ~~~~~~4$^{\rm c}$\\
IR-identified star clusters 		& $0.66 \pm 0.07$   	& &&		    & 				& Entire Galaxy     		      		    & ~~~~~~5\\  
Ultra-compact HII regions 		& 			&&& &	    			& Entire Galaxy					    & ~~~~~~6\\
Ultra-compact HII regions 		& 			&&& 31 	    &				&$-10\degree < \ell < 40\degree; \absb \le 0.5\degree$ & ~~~~~~7 \\
Ultra-compact HII regions		& 0.6		& && 30	    &				& Entire Galaxy	      		       	    & ~~~~~~8 \\			
HII regions				&			&&& 42	    & $-11$ & $17.9 < \ell < 55.4\degree; \absb < 1\degree$         & ~~~~~~9 \\
HII regions				&			&&& 39.3		    & $-$7.3  & Entire Galaxy, $\rgal < R_0$	  		    & ~~~~~10$^{\rm d}$ \\
HII regions				&			&&& $33.0\pm0.06$		    & $-$7.6  & Entire Galaxy, $\rgal < R_0$	  		    & ~~~~~11$^{\rm e}$ \\
High mass star forming regions 		&		    	&&& $29\pm0.5$   & $-20$ to 0 & $17.9 < \ell < 55.4\degree; \absb < 1\degree$    & ~~~~~12\\ 
\methanol\ masers 	     		& $0.4\pm0.1$	& &&  		    & 	       			& $-174\degree < \ell < 33\degree$     		    & ~~~~~13\\ 
HI cold neutral medium			&&&& $\sim 150$   &				& Entire Galaxy			    & ~~~~~14\\
CO					&$0.45$		&		    && &				& Entire Galaxy				    & ~~~~~15\\
Far-IR dust				&$0.50$		&		    && &				& Entire Galaxy				    & ~~~~~16\\
Far-IR dust				&$0.32$		&$-0.06$		&& && $-70\degree < \ell < 68\degree; \absb < 1.0\degree$				    & ~~~~~17\\
158 \,\micron\ [CII]			&$0.56$&&& 73	    & $-28$ & Entire Galaxy at $b=0\degree$			    & ~~~~~18\\ 
\hline

    \end{tabular}
  \end{centering}
  \\$^{\rm a}${As listed in the paper, regardless of whether value corresponds to the exponential or Gaussian scale height (see Equation~\ref{eq:scale_height}).}\\
  $^{\rm b}${1:\,\citet{beuther12}; 2:\,\citet{wienen15}; 3:\,\citet{rosolowsky10}; 4:\,\citet{ellsworth-bowers15}; 5:\,\citet{mercer05}; 6:\,\citet{bronfman00}; 7:\,\citet{becker94}; 8:\,\citet{wc89a}; 9:\,\citet{anderson09a}; 10:\,\citet{paladini04}; 11:\,\citet{bobylev16}; 12:\,\citet{urquhart11}; 13:\,\citet{walsh97}; 14:\,\citet{kalberla03}; 15:\,\citet{dame87}; 16:\,\citet{beichman88}; 17:\,\citet{molinari15}; 18:\,\citet{langer14}}\\
  $^{\rm c}${These authors correct for the solar offset
      above the Galactic midplane.  Quoted values are our computations
      from their database, using $z = d_\odot \sin (b)$ and the \citet{brand86}
      rotation curve.}\\
  $^{\rm d}${Values are only for the sample with the most accurate distances.}\\
  $^{\rm e}${Listed scale height is from the authors' Gaussian model fit.}
  \end{scriptsize}
\end{table*}

Previous studies of the vertical distribution of objects and solar height above the plane can be categorized as either using stellar or gas samples.  Solar height studies are summarized in \citet{humpfreys95} and \citet{karim17}.  Studies of stellar samples have a long history;
perhaps the first such study was done by \citet{vantulder42}, who
found an asymmetry in the stellar distribution that implied that the
Sun is $14\pm2$\,pc above the plane.  
Typical stellar studies examine discrepancies in the number of sources toward the north and south Galactic poles to determine the solar height \citep[e.g.,][]{humpfreys95}.  A typical value for the solar height from stellar studies is 20\,pc;
for example, \citet{maiz-apellaniz01} used OB
stars from $Hipparchos$ to derive $z_\odot = 24.2\pm1.7$\,pc,
\citet{chen01} used stars from an early release of the Sloan Digital
Sky Survey (SDSS) to derive $z_\odot = 27\pm4$\,pc., and
\citet{juric08} found using SDSS data release 3 (with some data
release 4) that the $z$-distributions for stars of a range of colors
and brightnesses are all consistent with $z_\odot \simeq
25$\,pc.  

We summarize the studies of Galactic latitude and $z$-distributions that use gas tracers in Table~\ref{tab:previous}, focusing on
works that use tracers sensitive to HMSF.  This
table contains the peak and scale height of the distributions.  If the
fits were exponential, we list the stated scale height.  If the fits
were Gaussians, we list the scale height $h$ as \begin{equation}
    h = \frac{\rm FWHM}{2(2\ln 2)^{0.5}}\,,
    \label{eq:scale_height}
\end{equation}
where FWHM is the full width at half maximum of the distribution.  For a given sample, we
do not expect significant discrepancies between the exponential and
Gaussian scale heights \citep{bobylev16}.  There are larger
discrepancies between values derived using gas tracers compared to
those derived using stellar tracers.  There is, nevertheless, good
agreement that the various distributions peak below the Galactic
midplane.  The tracers that are most sensitive to HMSF have narrow distributions with scale heights $\sim40\,\pc$,
whereas the distributions of \hi, CO, far-infrared emitting dust, and
\cii\ are broader.  

The solar height above the plane derived using gas tracers is generally lower than that found from stellar tracers \citep[see compilation in][]{karim17}.  Typical values are near 10\,\pc.  For example, \citet{bobylev16} found $z_\odot = 8\pm2\,\pc$ from a sample of \hii\ regions, masers, and molecular clouds and \citet{paladini03} found $z_\odot = 9.3\pm2\,\pc$ using a sample of \hii\ regions.

Because it was defined using low-resolution data and our measurements have since improved significantly, the IAU-defined
Galactic midplane may need to be revised \citep{goodman14}.
We now know that \sgra\ lies at $b = -0.046165\arcdeg$
\citep{reid04}, which places it below the IAU-defined location of the
Galactic Center (although by the IAU's definition Sgr\,A is at the
Galactic center \citep{blaauw60}).  \citet{goodman14} investigated the
implications of this offset and of the Sun lying above the midplane using the extremely long ``Nessie''
infrared dark cloud (IRDC).  Although Nessie lies below the midplane
as it is currently defined, because of the Sun's offset and the offset
of \sgra\ from $b=0\degree$, they found that Nessie may actually lie
in what they call the ``true'' midplane, which is tilted by angle $\thetat$
with respect to the IAU midplane definition\footnote{We call this
  rotation the ``tilt'' angle to be consistent with previous authors,
  although by convention it would be called the ``pitch'' angle.}.  While
suggestive, this study needs to be expanded to a larger sample of
objects in order to make stronger claims about the midplane
definition.

Tracers of HMSF formation should define the Galactic
midplane, although it is difficult to create a large, unbiased sample of HMSF regions.  Most of the gas tracers are related to massive stars, which
are born in the most massive molecular clouds in the Galaxy.
The high-mass stars themselves have lifetimes short enough that they
are unable to travel far from their birthplaces.  For example, an
O-star with a space velocity of 10\,\kms\ can only travel 100\,pc out
of the midplane in 10\,Myr, and only then if its velocity is entirely
in the ${\hat z}$ direction.  Other tracers of high-mass stars should
be similarly restricted to the midplane.



In Section~\ref{sec:define}, we first develop the methodology needed to redefine the Galactic midplane.  We apply this methodology to the {\it WISE} Catalog of Galactic \hii\ Regions
\citep{anderson14} in Section~\ref{sec:midplane}, after first characterizing the vertical structure of the
Galaxy's \hii\ region population in Section~\ref{sec:wise}.  We therefore define the HMSF midplane, determine the tilt and roll angles of the HMSF midplane with respect to the current IAU definition and determine the Sun's displacement from the
HMSF midplane.
The {\it WISE} catalog does not suffer from the same incompleteness and biases of other studies, and so may be better suited to determining the HMSF midplane than tracers used previously.


\section{Defining the Galactic Midplane \label{sec:define}}

Here, we develop the methodology required to define the midplane using a sample of discrete Galactic objects. Although the derived equations are general, we assume in later sections that the midplane passes through \sgra.  Future analyses with more data points may be able to relax this assumption.

\begin{figure}
  \begin{centering}
\includegraphics[width=3in]{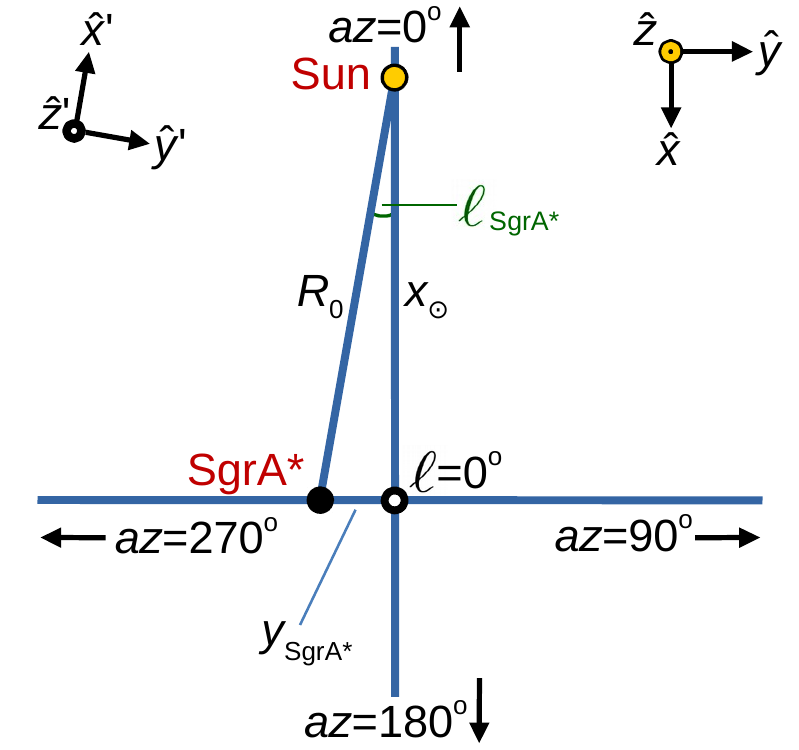}
\caption{Face-on geometry of the two coordinate systems used here, as
  viewed from the Galactic north pole (along $\hat{z}^\prime$).
  Angles are exaggerated for clarity.  Unprimed coordinates are
  centered on the Sun, whereas primed coordinates are centered at
  \sgra. Angles are indicated with green lines and font. Although $\hat{z}$ is not exactly out of the plane due to the
  midplane tilt, we ignore this complication when showing the
  coordinates. \label{fig:faceon}}
\end{centering}
\end{figure}

\subsection{Coordinate systems}
To define the Galactic midplane, we need to use two coordinate systems: the current IAU Galactic coordinate system
centered on the Sun ($x$, $y$, and $z$) and a new one
centered on \sgra\ $(x^{\prime}, y^{\prime}, z^{\prime})$ using the
``modified'' midplane definition.  
A Galactic azimuth ($az$) of zero degrees connects the Sun and the
Galactic Center, and azimuth increases clockwise in the plane as
viewed from the north Galactic pole.\footnote{Technically, $az$ is different between the two coordinate systems.  We use here the azimuth defined in the current coordinate system, but only to orient the reader.}  In the Sun-centered coordinates,
$\hat{x}$ points from the Sun to the (currently-defined) Galactic center, $\hat{y}$ points
in the direction of Galactic azimuth $az = 90\degree$, and $\hat{z}$
points toward the Galactic north pole.  In the \sgra-centered
coordinate system, $\hat{x}^\prime$ points from \sgra\ in the approximate direction of the Sun, $\hat{y}^\prime$ points in the direction of Galactic
azimuth $az \simeq 90\degree$, and $\hat{z}^\prime$ points approximately toward the
Galactic north pole.

We show the geometries of the two coordinate systems in Figures~\ref{fig:faceon},
\ref{fig:geometry}, and \ref{fig:geometry_roll}.
The modified midplane can be tilted by angle \thetat\ (rotated
about $\hat{x^\prime}$; Figure~\ref{fig:geometry}) and rolled by angle \thetar\ (rotated
about $\hat{y^\prime}$; Figure~\ref{fig:geometry_roll}).  
The modified midplane takes the form
\begin{equation}
    z^\prime = \thetat (x - x_{\sgramath}) + \thetar(y - y_{\sgramath}) + (z-z_{\sgramath})\,.
    \label{eq:plane}
\end{equation}
\sgra\ is located at $(\ell_{\sgramath}, b_{\sgramath}) =\\
(359.944249\degree,-0.046165\degree)$ 
\citep{reid04}, which gives non-zero values for $y_{\sgramath}$ and $z_{\sgramath}$.
As can be seen in Figure~\ref{fig:geometry},
\begin{equation}
  z_{\sgramath} = R_0\sin b_{\sgramath}\,.
\end{equation}
We can use the geometries in Figures~\ref{fig:faceon} and \ref{fig:geometry_roll} to determine
\begin{equation}
  y_{\sgramath} = \frac{R_0\sin \ell_{\sgramath}}{\cos\thetar}\,.
\end{equation}

  \begin{table*}
    \centering
      \caption{Reference points in the two coordinate systems}
      \begin{tabular}{lccccccccccccc}
        \hline
        Location                                   & Sun-centered                               & \sgra-centered             \\
        $(\ell, b, d)$                             & $z$                                        & $z^\prime$                  \\\hline 
        Sun: (---,---,0)                           & $0$                                        & $R_0\sin\thetat\cos\thetar - z_{\sgramath}\cos\thetat\cos\thetar = z_\odot^\prime$  \\
        \sgra: $(\ell_{\sgramath}, b_{\sgramath}, R_0)$ & $R_0\sin b_{\sgramath} = z_{\sgramath}$         & $0$               \\
        Current GC: $(0\degree, 0\degree, R_0)$    & $0$                                        & $-z_{\sgramath}\cos\thetat\cos\thetar$ \\
\hline
      \end{tabular}
  \label{tab:coords}
\end{table*}

We derive conversions between these coordinate systems in
Appendix~\ref{sec:appendix}, with the main result being the derivation
of $z^\prime$:
\begin{equation}
  \begin{split}
  z^\prime =&~(R_0-x)\sin\thetat\cos\thetar\\
  & - (y_{\sgramath}-y)\sin\thetar \\
  & + (z-z_{\sgramath})\cos\thetat\cos\thetar
  \end{split}
  \label{eq:zprime}
\end{equation}
where 
\begin{equation}
\left( \begin{array}{c} x \\ y \\ z  \end{array} \right) =
\left( \begin{array}{c} d_{_\sun} \cos \ell\ \cos b \\ d_{_\sun} \sin \ell\ \cos b \\ d_{_\sun} \sin b \end{array} \right)\,.
\label{eq:local_coord}
\end{equation}
We can therefore compute $z^\prime$ for each Galactic object, given its $(x,y,z)$ values, the rotation angles, and the
location of \sgra.
We give the
$z$-heights for locations along $az = 0\degree$ in
Table~\ref{tab:coords}.

\subsubsection{Midplane tilt, midplane roll, and the Sun's height}
\begin{figure*}
\begin{centering}
\includegraphics[width=6.0in]{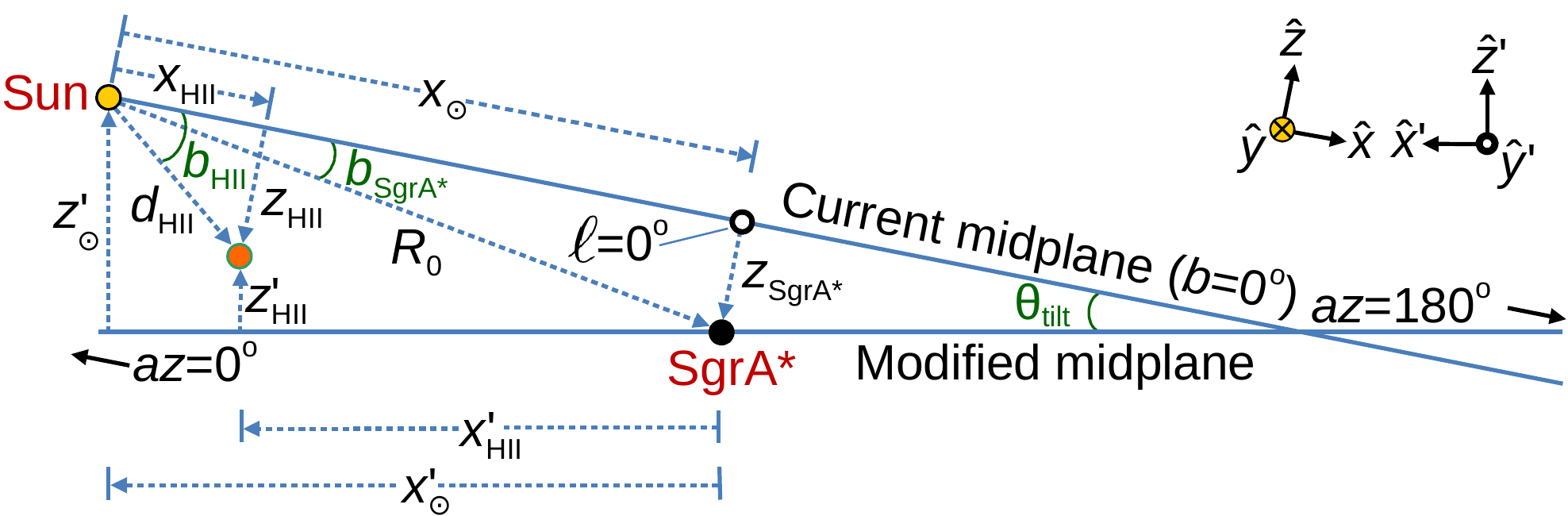}
\caption{Geometry used for calculations of the midplane tilt rotation
  angle and the Sun's height above the plane.  Angles are
  exaggerated for clarity and are indicated with green lines and font.
  Primed quantities correspond to distances in the coordinate system
  defined by the modified midplane that passes through \sgra.  The
  modified midplane is tilted by an angle $\thetat$, for a cut
  through the plane at Galactic azimuth of $0\degree$ as viewed from
  an azimuth of $270\degree$. 
  One example \hii\ region is shown to represent the analysis done in the next section on the entire
  \hii\ region population.  Although $\hat{y}$ is not
  exactly out of the plane due to the fact that $\ell_{\sgramath} \ne
  0$, we ignore this complication when showing the coordinates.\label{fig:geometry}}
\end{centering}
\end{figure*}

The tilt angle, which is apparent in Figure~\ref{fig:geometry}, does not have a compact analytical form unless we make
some simplifying assumptions.  Its complete form can be found by
solving (cf. Appendix~\ref{sec:appendix}):
\begin{eqnarray*}
  z_\odot^\prime = R_0\sin\thetat\cos\thetar - y_{\sgramath}\sin\thetar\\ + z_{\sgramath}\cos\thetat\cos\thetar\,.
\end{eqnarray*}
To simplify the equation for \thetat, we assume that 
$\theta_{\rm tilt} \simeq 0$ \citep[][find $\thetat \simeq 0.1\degree$]{goodman14}, so
$\cos\thetat\simeq1$.  We can further assume that $\thetar\simeq 0$ so that $\cos \thetar \simeq 1$ and $y_{\sgramath} \sin \thetar$ is small compared to the other terms. The
tilt angle is then:
\begin{equation}
  \thetat \simeq \sin^{-1}\left(\frac{z^\prime_\odot+z_{\sgramath}}{R_0}\right)
  \label{eq:tilt}
\end{equation}
This differs from the angle used in \citet{ellsworth-bowers13} by the
additional term $z_{\sgramath}$.  

\begin{figure*}
\begin{centering}
\includegraphics[width=5.3in]{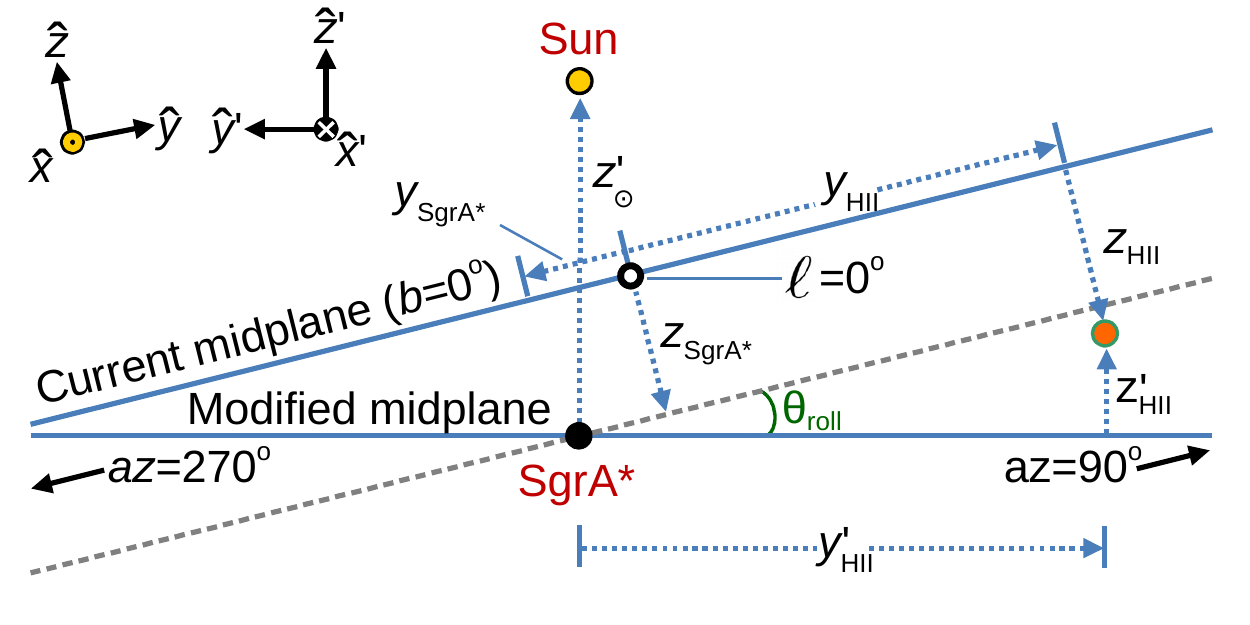}
\caption{Geometry used for calculations of the roll angle.  Angles are
  exaggerated for clarity and are indicated with green lines and font. Primed quantities correspond to distances
  in the coordinate system defined by the modified midplane that
  passes through Sgr~A$^*$.  We show the geometry for a roll angle
  $\thetar$, for a cut through the plane at Galactic azimuth of
  $90\degree$ as viewed from an azimuth of $180\degree$ (\sgra\ is in
  the foreground).  The third and fourth Galactic quadrants are therefore on the left of the diagram and the first and second quadrants are on the right. One example object, an \hii\ region, is shown. Although $\hat{x}$ is not exactly out of the plane due
  to the fact that $\ell_{\sgramath} \ne 0$, we ignore this
  complication when showing the
  coordinates. \label{fig:geometry_roll}}
\end{centering}
\end{figure*}

The roll angle is apparent in
Figure~\ref{fig:geometry_roll}.  
There is no compact solution for \thetar\ under reasonable assumptions.  Its full form can be found by solving Equation~\ref{eq:xyz_gal2}.


\begin{figure}
\begin{centering}
  \includegraphics[width=3.2in]{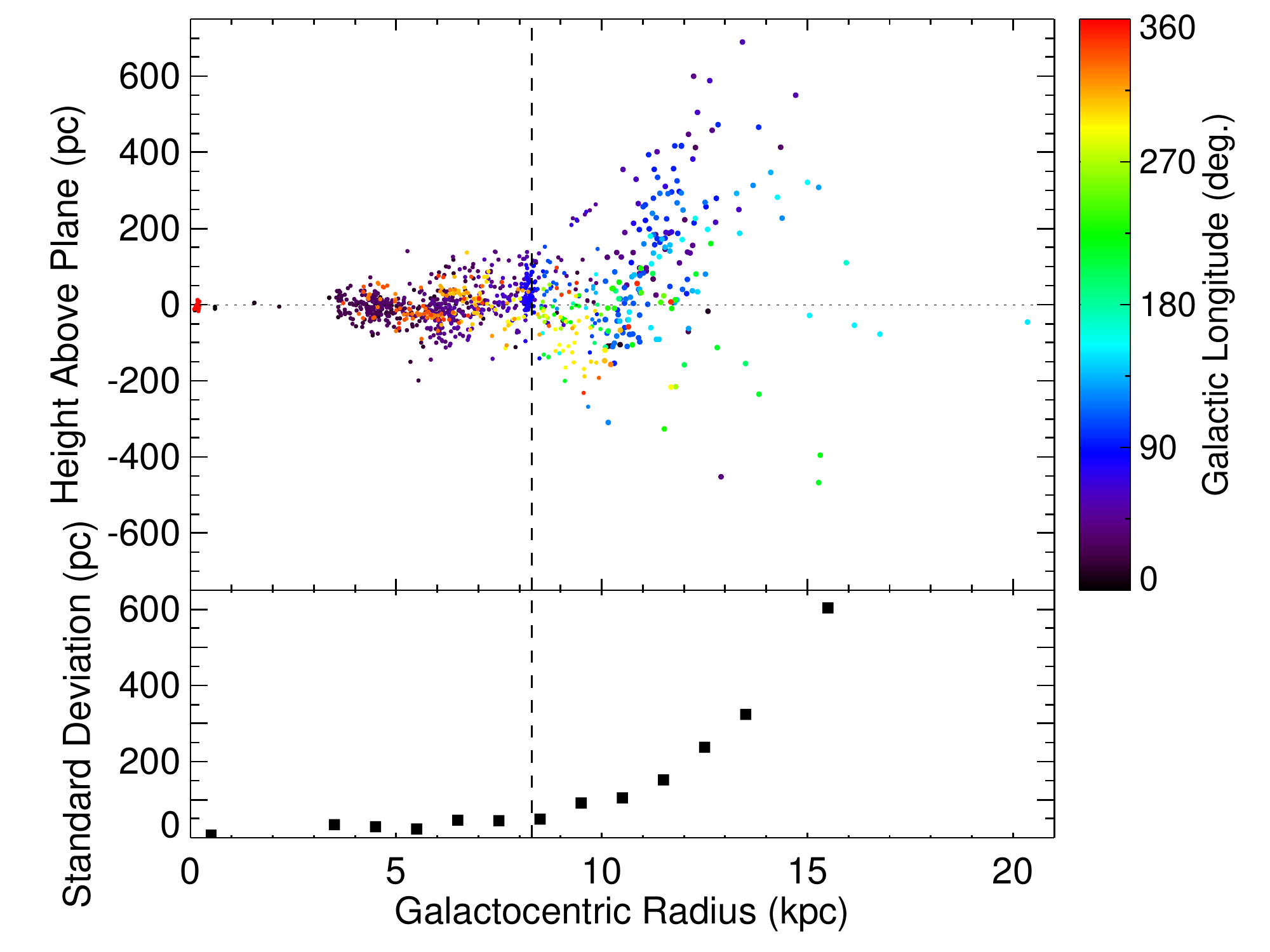}
\caption{The warping and flaring of the Galactic disk.  Shown is the
  height above the plane relative to IAU definition, $z$, versus
  Galactocentric radius \rgal\ for the \hii\ regions in the sample (top panel).  The color of each point
  corresponds to the Galactic longitude of the region.  We decrease
  the symbol size at low values of \rgal\ for clarity.  The
  warp begins near the solar circle, $\rgal \simeq 8.34\,\kpc$ (dashed
  vertical line), in agreement with previous studies.  The warp is
  toward the North Galactic pole in the first and second quadrants
  (black/purple/blue points) and toward the South Galactic pole in the
  third and fourth quadrants (green/yellow/orange points).  In the bottom panel, black
  filled squares show the standard deviation of the $z$-heights derived from Gaussian
  fits to sources in 1\,kpc\ bins.  Within the solar circle, the standard deviation is $\lsim 50\,\pc$, and this can be thought of as the scale height.  Outside the Solar circle, the standard deviation increases rapidly due to the Galactic warp.\label{fig:warp}}
\end{centering}
\end{figure}
\begin{figure*}
  \centering
  \includegraphics[width=6in]{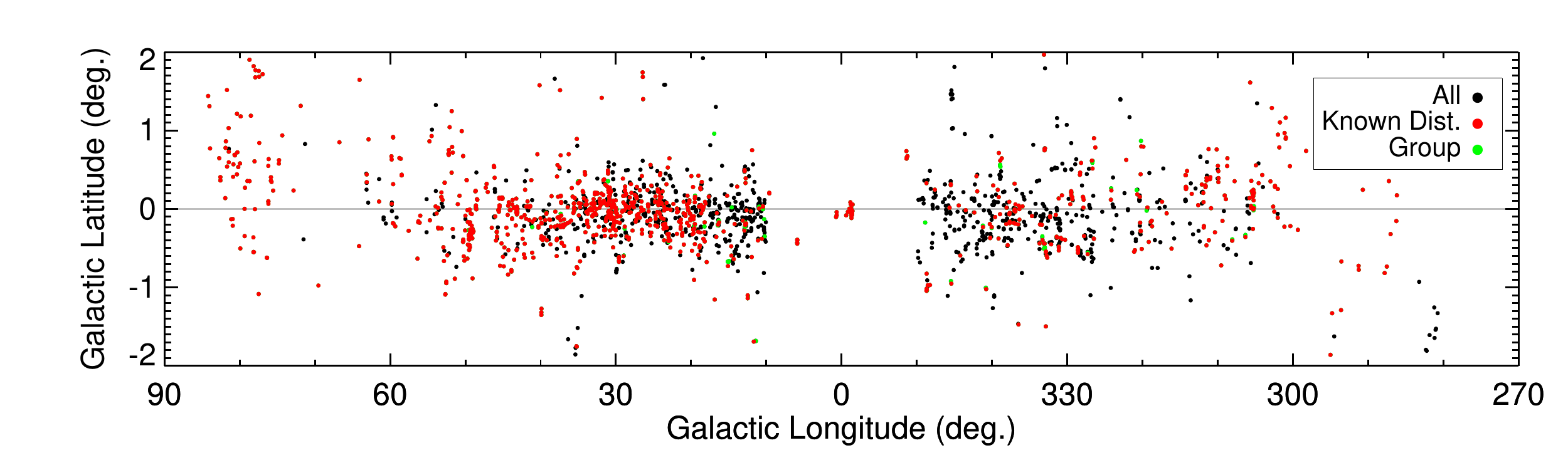}
\caption{Galactic distribution of \hii\ regions with $\rgal <
  R_0$.  Regions without known distances are shown as black dots,
  those with kinematic distances are shown with red dots, and the few
  ``group'' regions that are in large \hii\ region complexes but which lack individual spectroscopic observations are shown with
  green dots.  Unless they have trigonometric parallax distances or
  velocities consistent with the nuclear disk, sources within
  $10\degree$ of the Galactic Center lack known distances.  The
  latitude range here is restricted to show greater detail, and this
  excludes some regions from the plot.\label{fig:gal}}
\end{figure*}

\begin{figure}[!]
\begin{centering}
  \includegraphics[width=3in]{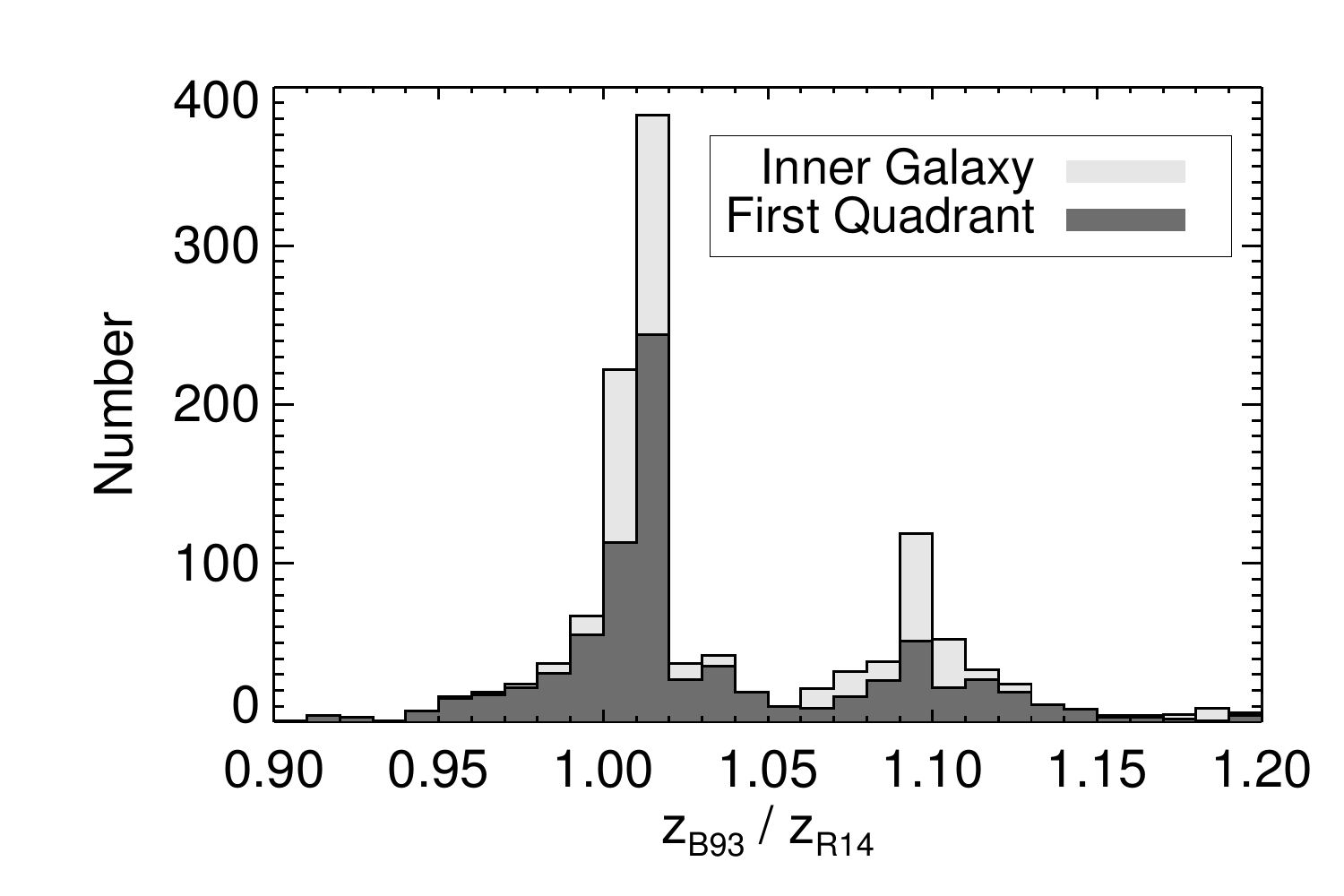}
  \includegraphics[width=3in]{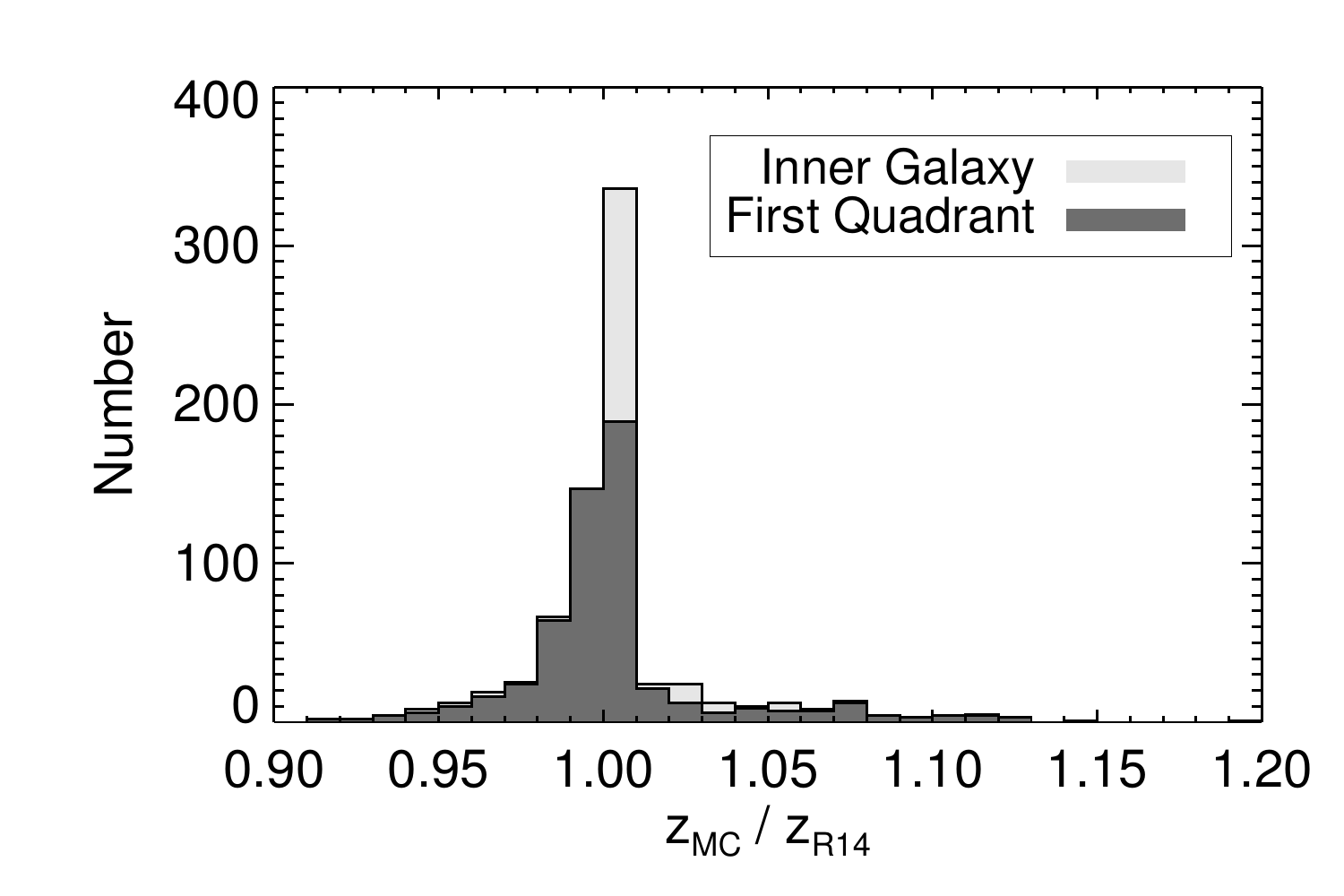}
\caption{Distribution of $z$-heights for the B93, R14, and MC distances, expressed as a ratio.  The B93 curve returns larger
  Heliocentric distances on average when compared with distances from the R14 curve and MC distances,
  and therefore on average has larger $z$ heights.  There is no
  significant difference between the inner Galaxy and first quadrant
  samples.  The bi-modal distribution in the top panel is due to B93 distances being preferentially larger than R14 distances for a given source velocity if the source is nearby, but the two rotation curves giving similar distances otherwise.  The larger uncertainties $z_{\rm B93}/z_{\rm R14} >1.05$ are all from sources at the near kinematic distance. This, combined with the fact that the distances themselves are not evenly distributed gives rise to the bi-modal nature.\label{fig:brand_reid}}
  \end{centering}
\end{figure}
\section{The {\it WISE} Catalog of Galactic HII Regions \label{sec:wise}}
We wish to investigate the HMSF midplane using
\hii\ regions in
the {\it WISE} Catalog of Galactic \hii\ Regions \citep[hereafter the {\it WISE} catalog][]{anderson14}, which contains all known Galactic
\hii\ regions.  First, however, we characterize this sample of \hii\ regions, and define subsamples to investigate how results may change when smaller numbers of \hii\ regions are used.
We use V2.1 of the {\it WISE}
catalog\footnote{http://astro.phys.wvu.edu/wise/}, which contains 1813
known \hii\ regions that have ionized gas spectroscopic observations,
1130 of which have known distances.  
This \hii\ region sample extends across the entire Galactic disk to
Heliocentric distances $>20$\,kpc and Galactocentric distances
$>18$\,kpc \citep{anderson15c}.  Since the {\it WISE} catalog was derived
using $6\arcsec$-resolution 12\,\micron\ data, or $2\arcsec$ {\it
  Spitzer} data in crowded fields, and the nominal \hii\ region size is on the order of arcminutes, confusion is minimal.  Therefore,
the {\it WISE} catalog suffers less from blending of distant regions
compared with lower resolution studies \citep[see][]{beuther12}.  The
catalog also has no latitude restriction, which
removes an
additional impediment to the study of the vertical distribution of
HMSF (see Section~\ref{sec:latitude}).

We compute the height above the plane, $z$, for each {\it WISE}
catalog \hii\ region using Equation~\ref{eq:local_coord}:
\begin{equation}
  z = d_\odot\, \sin (b)\,,
  \label{eq:z}
\end{equation}
where $d_\odot$ is the Heliocentric distance and $b$ 
latitude from the nominal \hii\ region centroid position in the
catalog.  The definition of $z$ has no correction for the Sun's height
above the midplane, and so differs from that used in some recent
studies \citep[e.g.,][]{ellsworth-bowers15}.  There is also no
correction for the displacement of \sgra\ below $b=0\degree$.

If available, the catalog distances are from maser parallax measurements \citep[e.g.,][]{reid09b, reid14}, but otherwise they are
kinematic distances.
The original {\it WISE} catalog used the
\citet[][hereafter B93]{brand93} rotation curve for kinematic
distances.  Here, we update all known \hii\ region distances using the method of \citet[][hereafer ``MC'']{wenger18}, which better accounts for uncertainties in kinematic distances.
Because of their large uncertainties, the catalog contains no kinematic
distances for \hii\ regions within $10\degree$ in Galactic longitude of the Galactic Center,
within $20\degree$ of the Galactic anti-center, and for any region
where the distance uncertainty is $>50\%$.
We use $R_0 = 8.34\,\kpc$ throughout \citep{reid14}.

Because of the Galactic warp, we cannot use all cataloged
\hii\ regions to investigate vertical structure in the Galaxy.  The
warp is known to begin around the solar orbit \citep{clemens88b}, at $R_0 \simeq
8.34$\,kpc.  We investigate the warp by plotting the $z$ distribution of
\hii\ regions as a function of \rgal\ in the top panel of Figure~\ref{fig:warp}.  
Each point in the top panel of Figure~\ref{fig:warp} represents an \hii\ region, color-coded by its Galactic longitude. In agreement with
previous results, the warp as traced by \hii\ regions begins near the
solar circle ($\rgal \simeq 8.34\,\kpc$) and
extends toward the north Galactic pole in the first Galactic quadrant
and toward the south Galactic pole in the third Galactic quadrant.  

The standard deviation of the \hii\ region sample shown in the bottom panel of Figure~\ref{fig:warp} is relatively constant in the inner Galaxy.  The inner Galaxy values are all $<50\,\pc,$ and this can be thought of as the scale height.  In the outer Galaxy, the standard deviation
increases with Galactocentric radius as a result of the Galactic warp.  This agrees with the results of
the \hii\ region study by \citet{paladini04}  and the CO study by
  \citet{malhotra94}.  
We exclude \rgal\ bins that have fewer than
10 sources from these computations.

In the analysis of the HMSF midplane (Section~\ref{sec:midplane}), we
exclude sources with $\rgal > R_0$, where $R_0=8.34\,\kpc$.  We also exclude regions with distance
uncertainties $>50\%$.
We show the Galactic locations of the \hii\ regions
studied here in Figure~\ref{fig:gal}.  The lack of regions within
$10\degree$ of the Galactic Center is in part caused by a lack of
known distances for those regions.  Most regions in the large
concentration near $\ell = 80\degree$ are associated with the Cygnus~X
complex.

\subsection{Subsamples of the {\it WISE} Catalog}
Due to issues of completeness and biases introduced by large
\hii\ region complexes, we define multiple subsamples of the {\it WISE}
catalog.  We run our analyses on these subsamples to investigate
potential biases in our results.  We also test the impact of using
different rotation curve models.  For all subsamples, we include only
regions with $\rgal < R_0$, with $R_0 = 8.34\,\kpc$.

\subsubsection{Galactic Quadrants}
The completeness of the {\it WISE} catalog varies across the Galaxy.
Nearly all recent \hii\ region surveys have taken place in the
northern sky, and therefore there are many more known \hii\ regions in
the first Galactic quadrant compared to the fourth.  The luminosity
distribution of the first quadrant sample suggests that it is complete
for all \hii\ regions ionized by single O-stars, but this is not the
case in the fourth quadrant (W.~Armentrout et al., 2018, in prep.).
This asymmetry may introduce a bias into our analysis.  We therefore
perform our analyses below using two Galactic longitude subsamples
that both have $\rgal < R_0$: one from $10\degree < \ell < 75\degree$
(hereafter the ``first quadrant sample'') and one containing all
regions in the first and fourth quadrants (the ``inner Galaxy
sample'').  The first quadrant sample has 682 \hii\ regions, 458 of which have known distances, and the inner Galaxy sample has 1149 \hii\ regions, 613 of which have known distances.

\subsubsection{HII Region Complexes}
\hii\ regions are frequently found in large complexes containing many
individual \hii\ regions.  The {\it WISE} catalog lists entries for each
individual region in the complex and therefore the results of a
statistical study will differ based on whether the complex is
considered to be one \hii\ region or many.  There are $\sim 600$
objects in the {\it WISE} catalog that do not have ionized gas or
molecular spectroscopic observations, but are placed into a complex on
the basis of the appearance of the complex in mid-infrared and radio
continuum data (e.g., W49, W51, Sgr~B2, etc.).  The distance to these
regions are assumed to be that of the other complex members.  These
large complexes may bias our results because there are many regions in
the catalog at particular Galactic locations.  This bias may be
warranted because these large complexes may better define the midplane (as
found by V.~Cunningham et al., 2018, in prep.).

We test for the effect of complexes on our results by running the
analyses on two subsamples, one only containing ``unique''
\hii\ regions (i.e., each complex contains only one catalog entry),
and the other that has all regions, including ``group'' regions that
 that are in large \hii\ region complexes but which lack individual spectroscopic observations.  For the group regions, we
assume the kinematic distance of the other complex members.  We only show results from these subsamples in the first Galactic quadrant.  In the first Galactic quadrant, the unique subsample contains 605 \hii\ regions, 408 of which have known distances, and the group subsample contains 1132 \hii\ regions, 725 of which have known distances.

\subsubsection{Rotation Curves}
The majority of the {\it WISE} catalog distances are kinematic, and
are therefore sensitive to the choice of rotation curve.  Different
distances result in different values for $z$
(cf.~Equation~\ref{eq:z}).  We examine how our results change when kinematic distances
are computed using the B93 curve, the \citet[][hereafter ``R14'']{reid14}, and the MC analysis.  We do not change the parallax distances in any of our trials.  R14 lists multiple
rotation curve models; the one we use here has a solar circular
angular velocity $\Omega_0 = 235\,\kms$, a solar distance from the
Galactic Center $R_0 = 8.34\,\kpc$, and $\Omega (\rgal) = \Omega_0 -
0.1\,\rgal$.  In general, the B93 curve gives larger distances
compared to the R14 curve, and therefore the B93 $z$- distances are
larger than those of R14 (Figure~\ref{fig:brand_reid}, top panel).  The MC and R14 curve $z$-distances are similar (Figure~\ref{fig:brand_reid}, bottom panel).

\begin{figure*}
\centering
  \includegraphics[width=3.3in]{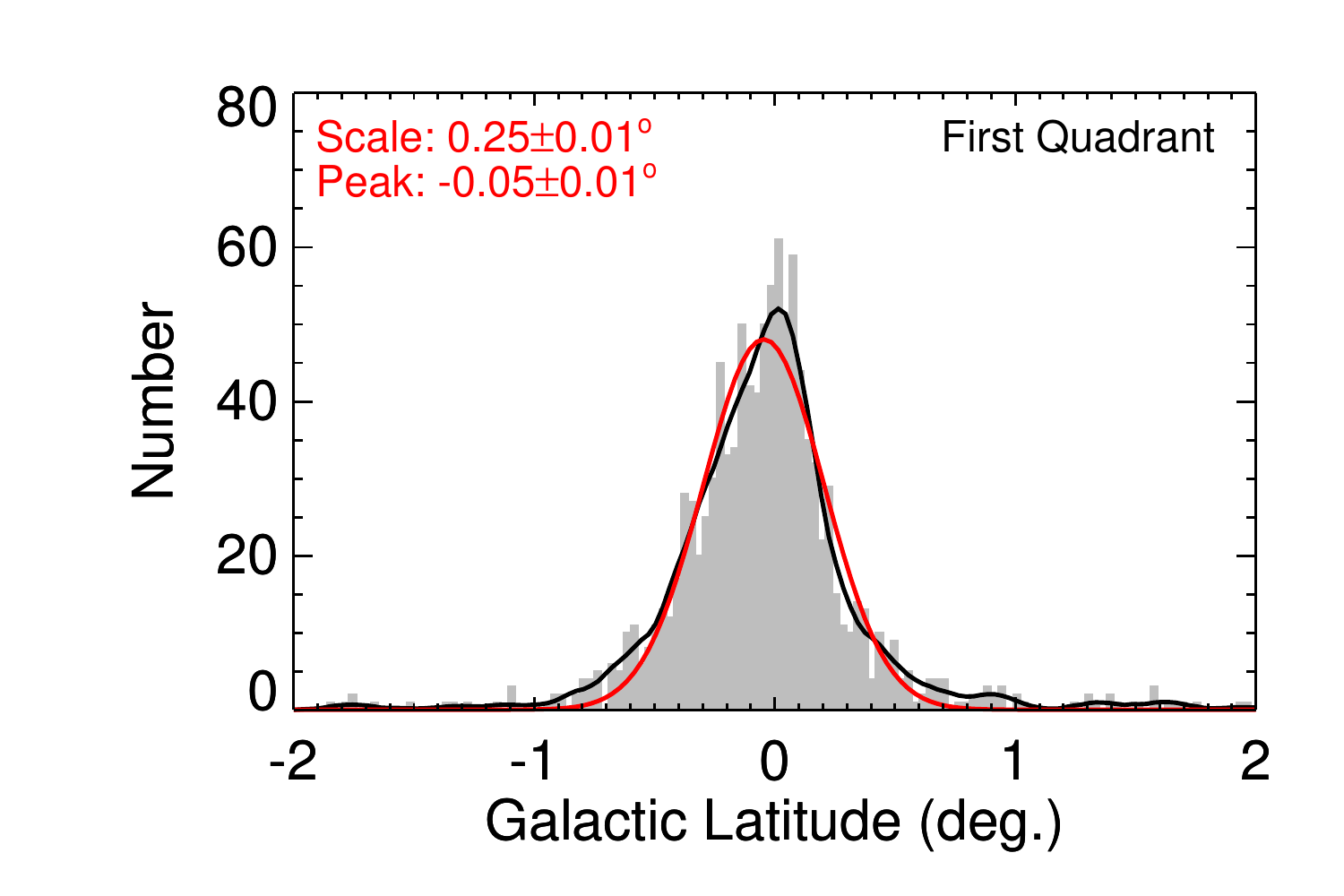}
\includegraphics[width=3.3in]{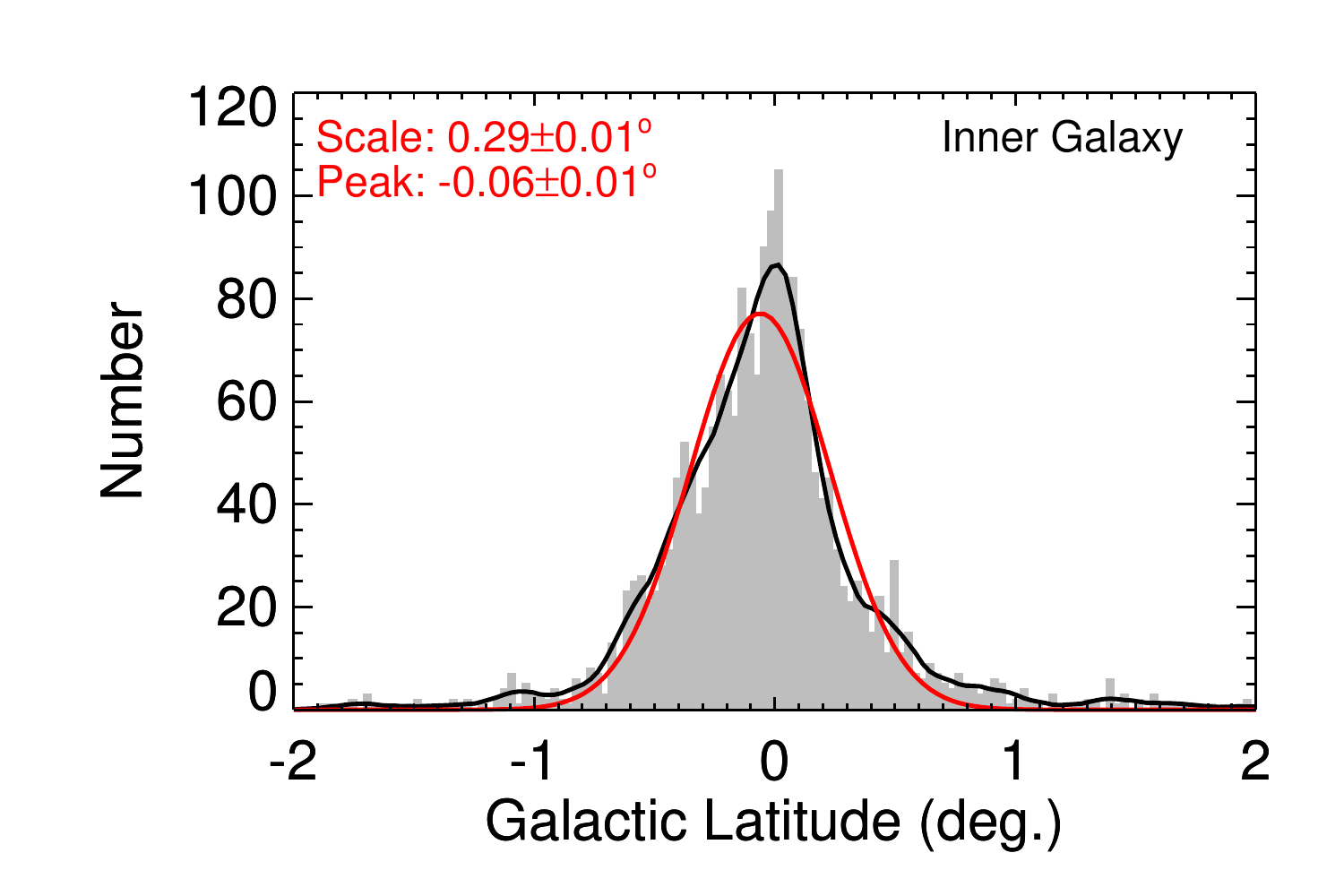}
\caption{Galactic latitude distributions for \hii\ regions in the
  first quadrant (left) and the inner Galaxy (right).
  The KDE is shown with a solid black curve, and this is fit with a
  Gaussian function shown as the red curve.  The Gaussian fits to all
  subsamples peak at small negative values of $b$.  That these
  distributions peak at negative latitudes can be explained if the Sun
  lies above the HMSF midplane.  The scale heights from the
  Gaussian fits are all near $0.30\degree$.}
\label{fig:bdist}
\end{figure*}

\begin{figure*}
  \centering
\includegraphics[width=3.3in]{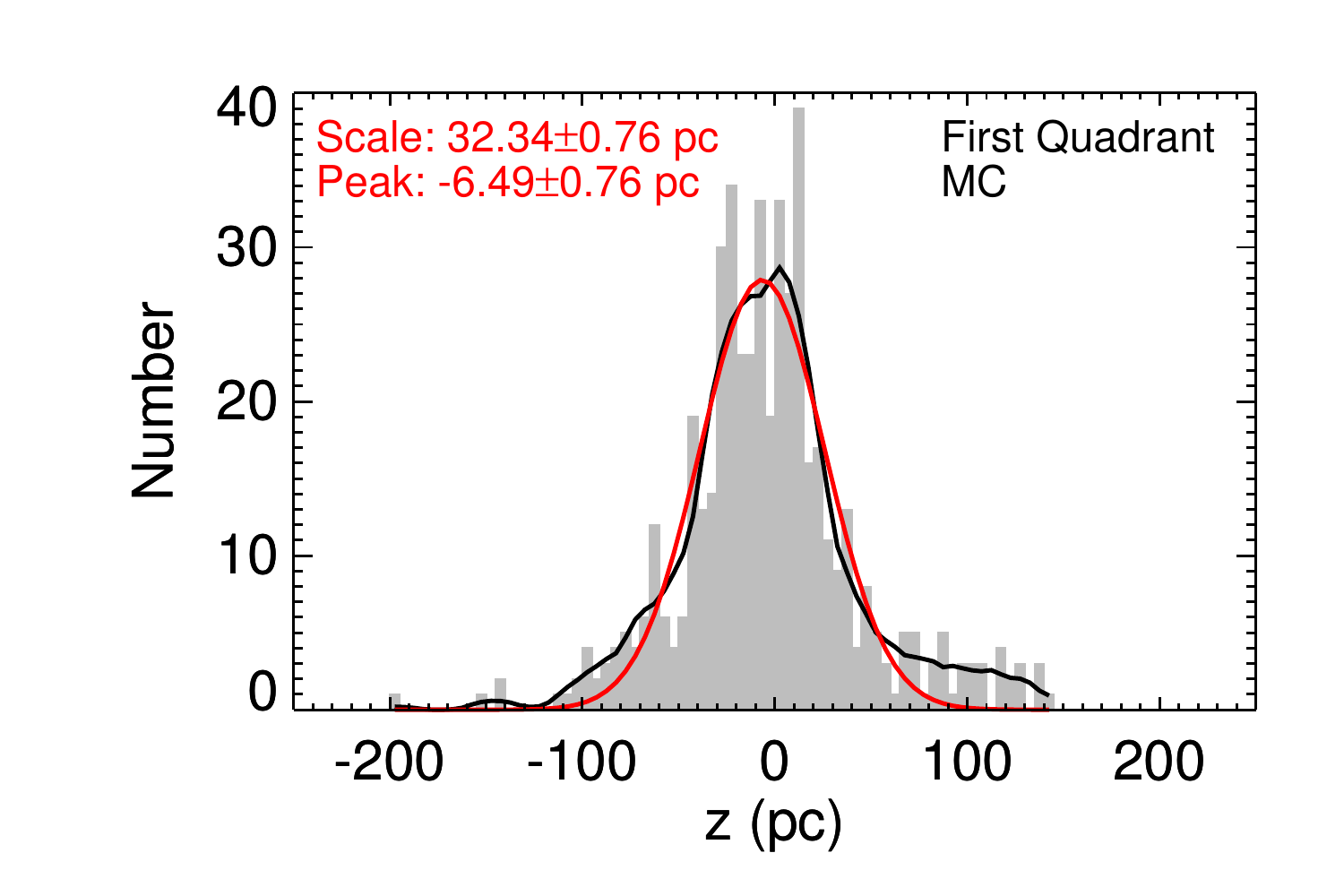}
\includegraphics[width=3.3in]{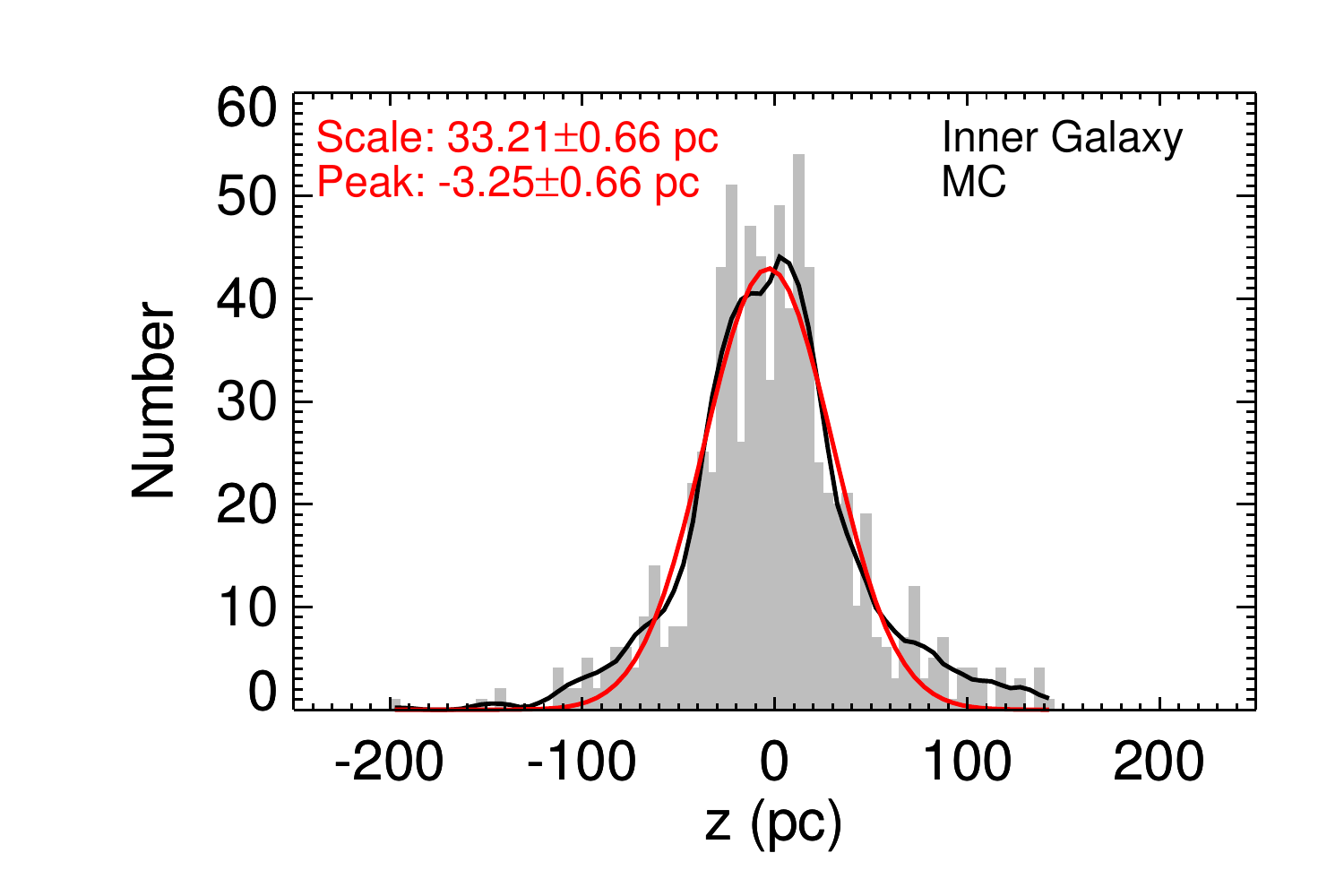}
\caption{Height above the plane, $z$, for the first quadrant (left) and inner Galaxy samples (right) for MC distances. The KDE is shown
  with a solid black curve, and this is fit with a Gaussian function
  shown as the red curve.  The Gaussian fits to the various subsamples
  peak at small negative values of $z$.  That these distributions peak
  at negative $z$-heights can be explained if the Sun lies above the
  HMSF midplane.\label{fig:zdist}}
\end{figure*}

\subsection{Characterizing the HII Region Vertical Distribution}
We characterize the {\it WISE} catalog Galactic latitude distribution for all \hii\ regions in the sample and the $z$-distribution for regions with known distances in Table~\ref{tab:bz}.  These analyses do not rely on the modified midplane definition in Section~\ref{sec:define}, but are comparable to those derived by previous authors in Table~\ref{tab:previous}.

\begin{table*}[]
  \centering
  \begin{footnotesize}
  \renewcommand\tabcolsep{2pt}
    \begin{centering}
      \caption{\hii\ Region $b$ and $z$-Distributions \label{tab:bz}}
      \begin{tabular}{lccccccccc}
        \hline
&&&\multicolumn{3}{c}{Galactic Latitude $b$} && \multicolumn{3}{c}{Height $z$}\\ \cline{4-6} \cline{8-10}
Sample & Rot. Curve & Modification & Number & Scale Height & Peak && Number & Scale Height & Peak \\
  &   &   &   & (deg.) & (deg.) & & & (pc) & (pc)\\\hline
First Quadrant & MC  &                  &682&$0.27\pm0.01$& $-0.05 \pm0.01$&&458&$32.3\pm0.8$& $-6.5\pm0.8$\\
First Quadrant & MC  & Unique           &605&$0.28\pm0.01$& $-0.05 \pm0.01$&&408&$33.5\pm0.8$& $-6.2\pm0.8$\\
First Quadrant & MC  & Group            &1132&$0.29\pm0.01$& $-0.05 \pm0.01$&&725&$32.3\pm0.8$& $-6.5\pm0.8$\\
First Quadrant & R14 &                  &682&$0.27\pm0.01$& $-0.05 \pm0.01$&&475&$30.1\pm0.8$& $-6.2\pm0.8$\\
First Quadrant & B93 &                  &682&$0.27\pm0.01$& $-0.05 \pm0.01$&&475&$30.8\pm0.8$& $-6.5\pm0.8$\\
Inner Galaxy   & MC  &                  &1149&$0.31\pm0.01$& $-0.05\pm0.01$&&613&$33.2\pm0.7$& $-3.2\pm0.7$\\
First Quadrant & MC  & $r < 2\,\pc$     &130&$0.21\pm0.01$& $-0.02 \pm0.01$&&104&$24.4\pm0.7$& $-4.6\pm0.7$\\
First Quadrant & MC  & $2 < r < 5\,\pc$ &211&$0.34\pm0.01$& $-0.04\pm0.01$&&200&$40.3\pm1.1$& $-6.0\pm1.1$\\
First Quadrant & MC  & $r > 5\,\pc$     &159&$0.29\pm0.01$& $-0.07\pm0.01$&&154&$41.2\pm0.9$& $-9.7\pm0.9$\\
        \hline
      \end{tabular}
    \end{centering}
  \end{footnotesize}
\end{table*}

\subsubsection{Galactic Latitude Distribution}
The distribution of Galactic latitudes is representative of
the $z$-height distribution, but since it does not require distances
to the objects the analysis can be done on a larger sample of
\hii\ regions.  Figure~\ref{fig:bdist} shows the first quadrant and inner Galaxy
\hii\ region Galactic latitude distributions, although we perform the same analysis for all subsamples defined previously, for all rotation curves.  We plot the ``kernel density estimation'' (KDE) in the
black curve.  The KDE estimates the underlying distribution, and an
analysis of the KDE is free from the uncertainties associated with the
choice of bin size.  For this and all subsequent analyses, we use the
``Epanechnikov'' kernel with the optimal bandwidth as suggested by \citet[][their Equation~3.31]{silverman86}. We fit the KDE distributions with
Gaussian functions and store the results in Table~\ref{tab:bz}.

For all subsamples, the peak of the Galactic latitude distribution is slightly below $b = 0\degree$
(possibly indicating a positive value for the solar height $z^\prime_\odot$), ranging from
$-0.04$ to $-0.06\degree$. The scale height, again the standard deviation of a Gaussian fit to the Galactic latitude distribution, is between $0.25\degree$
and $0.30\degree$ for all subsamples.
These scale height values are similar to those found for other
high-mass star tracers (Table~\ref{tab:previous}).  It is interesting
that our sample of \hii\ regions, which spans a wide range of
evolutionary stages, has the same scale height as tracers that are
more sensitive to future star formation (e.g., sub-mm/mm clumps).
We can infer that the lifetime of \hii\ regions is short enough to not make a
large difference in their $b$-distribution compared with younger
objects.

\subsubsection{$z$-Distribution}
For regions with known distances, we can study the
$z$-height distribution directly.
The $z$-distributions, for which we show examples in Figure~\ref{fig:zdist}, are
approximately Gaussian for all subsamples. There are, however,
``wings'' to the distributions at high and low values of $z$.  As with
the Galactic latitude distributions, we fit the KDEs of the $z$
distributions with Gaussian functions and store the results in Table~\ref{tab:bz}.
The analysis of the z-distribution is
necessarily limited to only \hii\ regions with known distances, which
is a smaller subsample compared to that used in the Galactic latitude
analysis.  All subsamples peak at small negative values, from $-3$ to $-6$\,\pc, again
implying that the Sun is located above the midplane.
The scale heights for the various subsamples range from $30$ to
$34\,\pc$. All distance methods return similar results.
These values are similar to those found for ultra-compact
\hii\ regions \citep{bronfman00}, sub-mm clumps \citep{wienen15} and
high-mass star forming regions \citep{urquhart11} (see
Table~\ref{tab:previous}).




\subsubsection{Variations with HII Region Size}
\begin{figure}[!]
  \centering
  \includegraphics[width=2.7in]{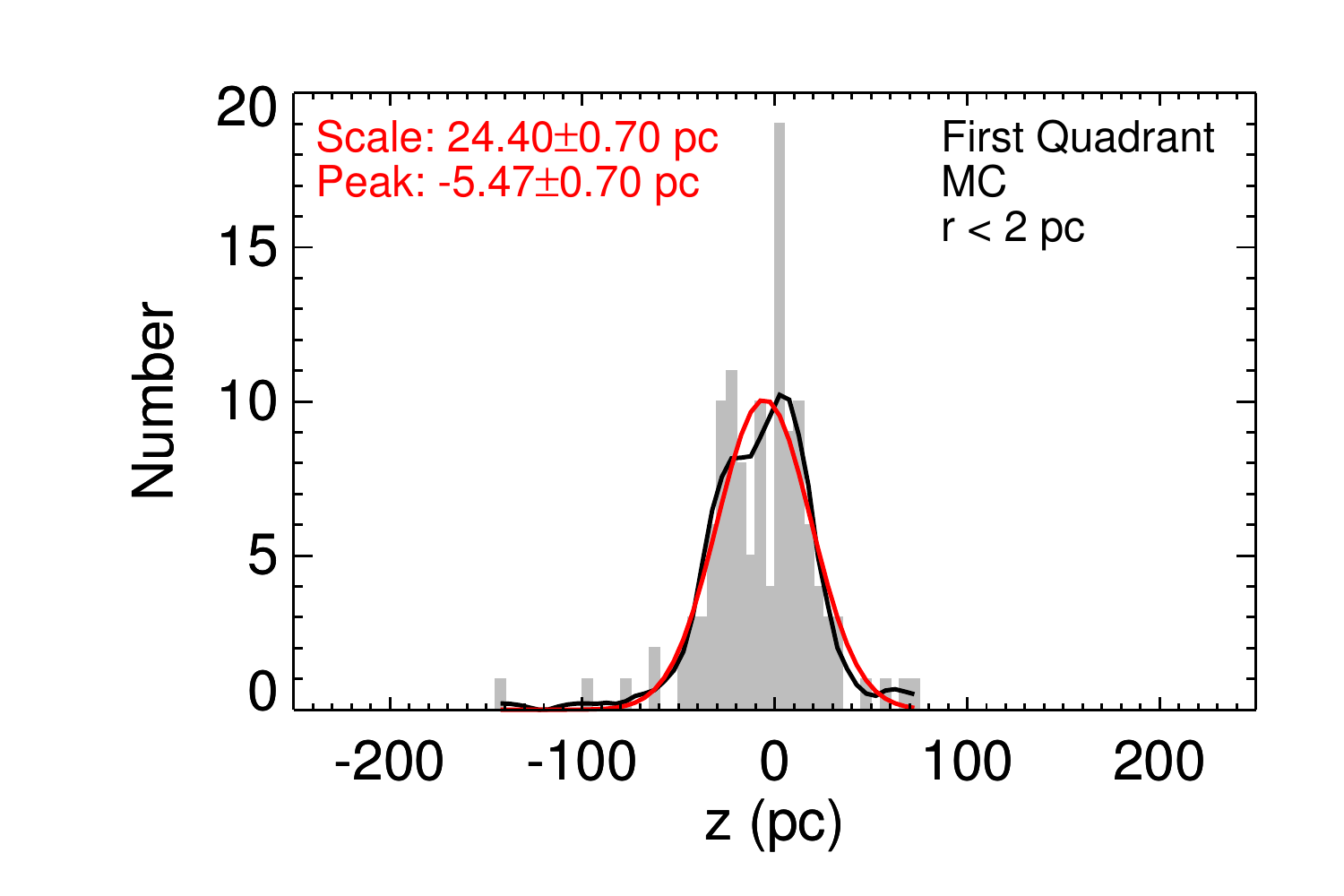}\vskip-15pt
  \includegraphics[width=2.7in]{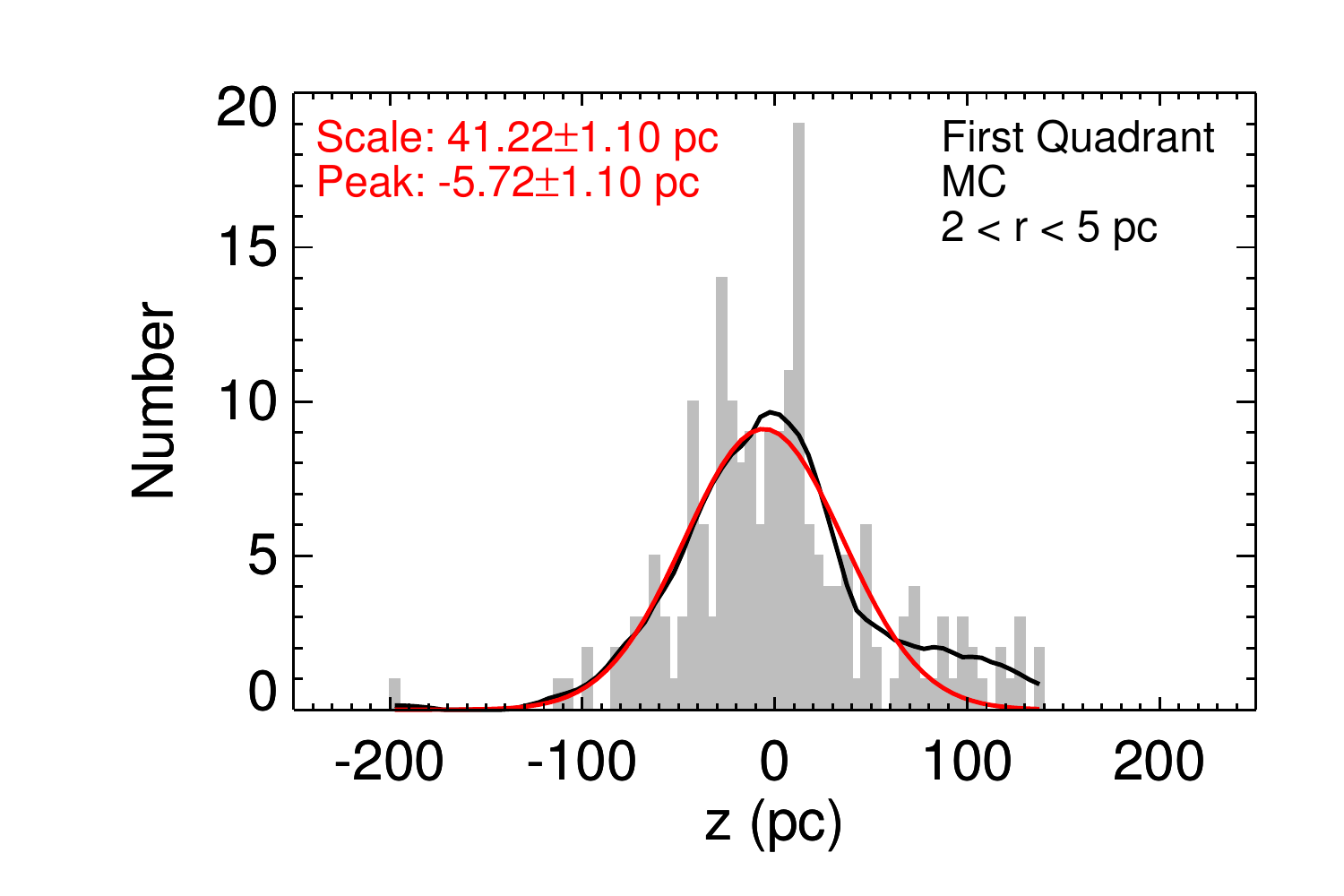}\vskip-15pt
  \includegraphics[width=2.7in]{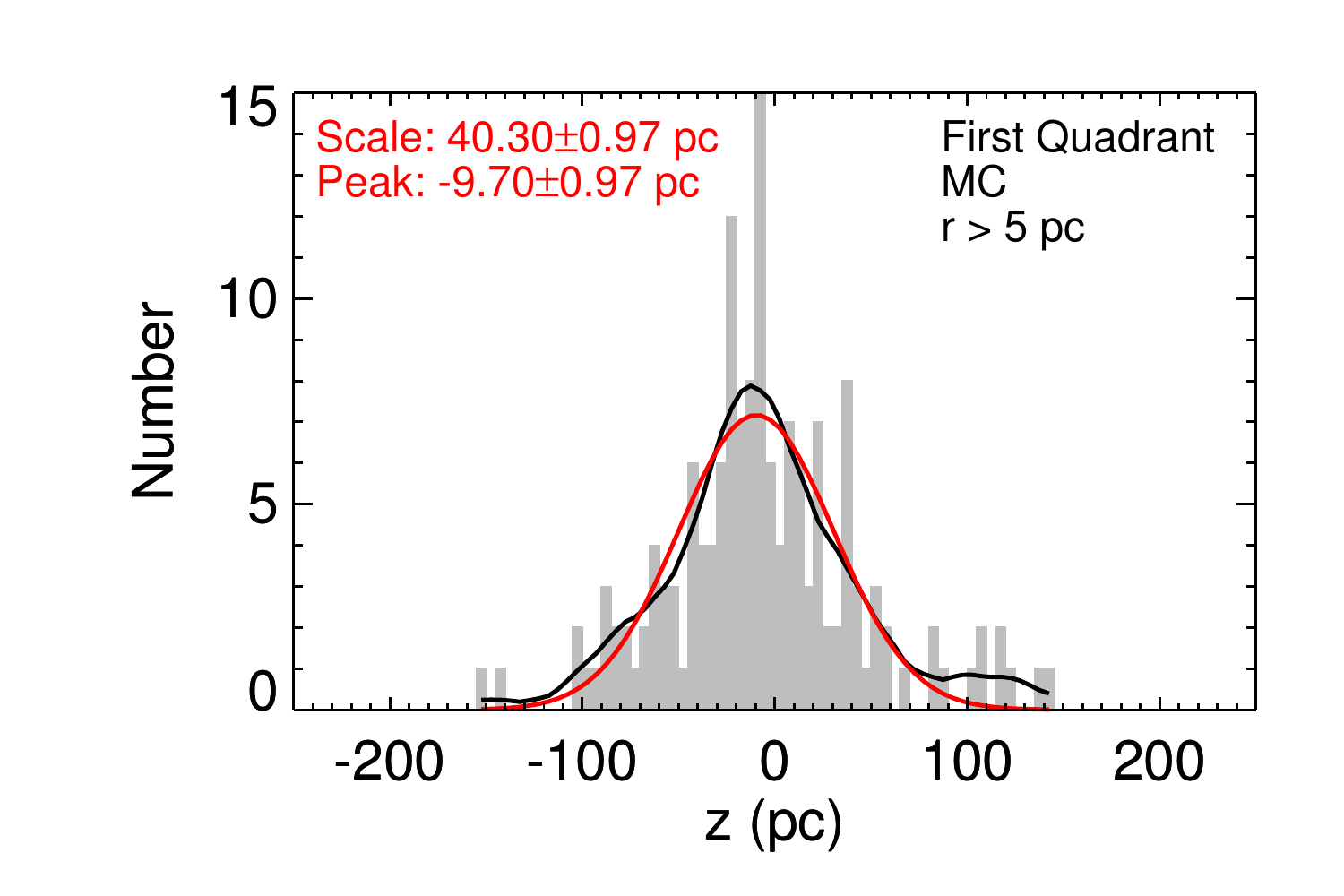}

  \caption{The $z$-distributions for small ($r < 2\,\pc$; top),
    medium ($2 < r < 5\,\pc$; middle), and large ($r > 5\,\pc$;
    bottom) first quadrant \hii\ region samples.  The smallest
    regions have the narrowest distribution.  The black lines are the
    KDEs and the red lines are Gaussian fits to the KDEs.  \label{fig:size}}
\end{figure}

Finally, we test how the $b$- and $z$-distributions changes when the sample is
restricted to \hii\ regions of various physical sizes.  
The size of an \hii\ region is a proxy for its age \citep[e.g.][]{spitzer78}.
Diffuse
\hii\ regions are difficult to detect \citep{lockman96, anderson17b},
and excluding larger diffuse regions from the sample may have an
impact on the derived results.  We divide the first quadrant sample
into three physical size groups based on the {\it WISE} catalog radius $r$:
$r < 2\,\pc$, $2 < r < 5\,\pc$, and $r > 5\,\pc$.  We show these
distributions and fits in Figure~\ref{fig:size}, and give the fit
parameters in Table~\ref{tab:bz}.  Smaller regions
have a smaller scale height of $\sim 25\,\pc$, whereas the largest regions
have scale heights of $\sim40\,\pc$.  Furthermore, the larger region distributions are consistent with larger solar heights and hence larger tilt angles.  Assuming the smaller regions are
on average younger, this result is consistent with migration of older
regions out of the plane as they age.

\begin{table*}
  \centering
  \renewcommand\tabcolsep{2pt}
  \begin{centering}
    \begin{footnotesize}
      \caption{HMSF Midplane Parameters\label{tab:fits}}
      \begin{tabular}{lccccCCCCCCCCC}
        \hline
& \multicolumn{3}{c}{} & \multicolumn{2}{c}{$\theta_{\rm roll} = 0\degree$} & & \multicolumn{3}{c}{$\theta_{\rm roll}$ Free} \\ \cline{5-6}\cline{8-10}
Sample & Rot. Curve & Modification & Number & $z_\odot^\prime$ & $\theta_{\rm tilt}$ &   & $z_\odot^\prime$ & $\theta_{\rm tilt}$ & $\theta_{\rm roll}$\\
  &   &   &   & (pc) & (deg.) &   & (pc) & (deg.) & (deg.)\\\hline
First Quadrant & MC  &         & 458  &  $5.5\pm2.5$ &-$0.01\pm0.01$ &   &     $5.6\pm0.7$ &$-0.01\pm0.01$  &   $0.08\pm0.01$\\
First Quadrant & MC  & Unique                 & 408  &   $4.5\pm2.5$ & $-$0.02\pm0.01$ &   &     6.5\pm0.8 &0.00\pm0.01  &   $0.10\pm0.01$\\
First Quadrant & MC  & Group                  & 725  &   $5.5\pm2.5$ & $-$0.01\pm0.01$ &   &     9.4\pm0.6 &   0.02\pm0.01  &   $0.11\pm0.01$\\
First Quadrant & R14 &                  				  & 475  &   $5.0\pm2.5$ & $-$0.01\pm0.01$ &   &     4.2\pm1.1 &$-$0.02\pm0.01  &   0.03\pm0.01\\
First Quadrant & B93 &                        & 475  &   $5.5\pm2.5$ & $-$0.01\pm0.01$ &   &     5.7\pm1.2 &  $-$ 0.01\pm0.01  &   0.04\pm0.01\\
Inner Galaxy   & MC  &                        & 613  &  $-2.8\pm1.6$ & $-$0.07\pm0.01$ &   &     1.3\pm0.4 &$-$0.04\pm0.01  &   0.04\pm0.01\\
First Quadrant & MC  & $d_{\rm cut} = 4.7\,\kpc, \alpha = -0.66$    & 306  &   $4.8\pm3.1$ & $-$0.01\pm0.02$ &   & \nodata & \nodata & \nodata\\
First Quadrant & MC  & $|b|<0.5\degree$ 	 	  	     	  & 385  &   $3.5\pm2.5$ & $-$0.02\pm0.02$ &   &    15.0\pm0.6 &    0.06\pm0.01 &   0.05\pm0.01\\
First Quadrant & MC  & $|b|<1.0\degree$       & 441  &   $6.2\pm1.6$ &    0.00\pm0.02$ &   &    14.3\pm0.7 &    0.05\pm0.01 &   0.09\pm0.01\\
First Quadrant & MC  & $r < 2\,\pc$           & 104  &   $0.8\pm1.6$ & $-$0.04\pm0.01$ &   &  $-$3.3\pm1.3 & $-$0.07\pm0.01 &$-$0.04\pm0.01\\
First Quadrant & MC  & $2 < r < 5\,\pc$       & 200  &   $3.5\pm3.0$ & $-$0.02\pm0.02$ &   &    18.0\pm1.2 &    0.08\pm0.01 &   0.15\pm0.01\\
First Quadrant & MC  & $r > 5\,\pc$           & 154  &  $18.5\pm3.5$ &    0.08\pm0.03$ &   &  $-$8.6\pm1.5 &    $-$0.11\pm0.01 &   0.02\pm0.01\\
        \hline
      \end{tabular}
    \end{footnotesize}
  \end{centering}
\end{table*}
\section{The HMSF Midplane Defined by HII Regions\label{sec:midplane}}
We use the results from Section~\ref{sec:define} to define the HMSF midplane with \hii\ regions.  In this process, we determine the tilt and roll angles, and also the solar offset from the midplane.
Our analysis necessarily ignores local deviations
\citep[e.g.,][]{malhotra94, malhotra95} to find the average midplane
definition most consistent with the data.  As before, we exclude \hii\ regions with $\rgal > R_0$, \hii\ regions within $10\degree$ in Galactic longitude of the Galactic Center, all regions with distance uncertainties $>50\%$.




\subsection{Midplane Tilt with $\thetar = 0\degree$\label{sec:tilt}}
We assume that the \hii\ region $z^\prime$ distribution will peak at 0\,\pc\ for the ``correct''
value of $z^\prime_\odot$.  
Changing $z^\prime_\odot$ alters
$z^{\prime}$ for each \hii\ region in the sample
(Equation~\ref{eq:zprime}). This value of $z^\prime_\odot$ also
results in a unique value for $\thetat$
(Equation~\ref{eq:tilt}).
With the limited number of fourth quadrant \hii\ regions with known distances,
$\thetar$ is difficult to constrain, and is often not acounted for in other analyses of the midplane \citep[e.g.,][]{goodman14}.  

\begin{figure*}
  \centering
  \includegraphics[width=3in]{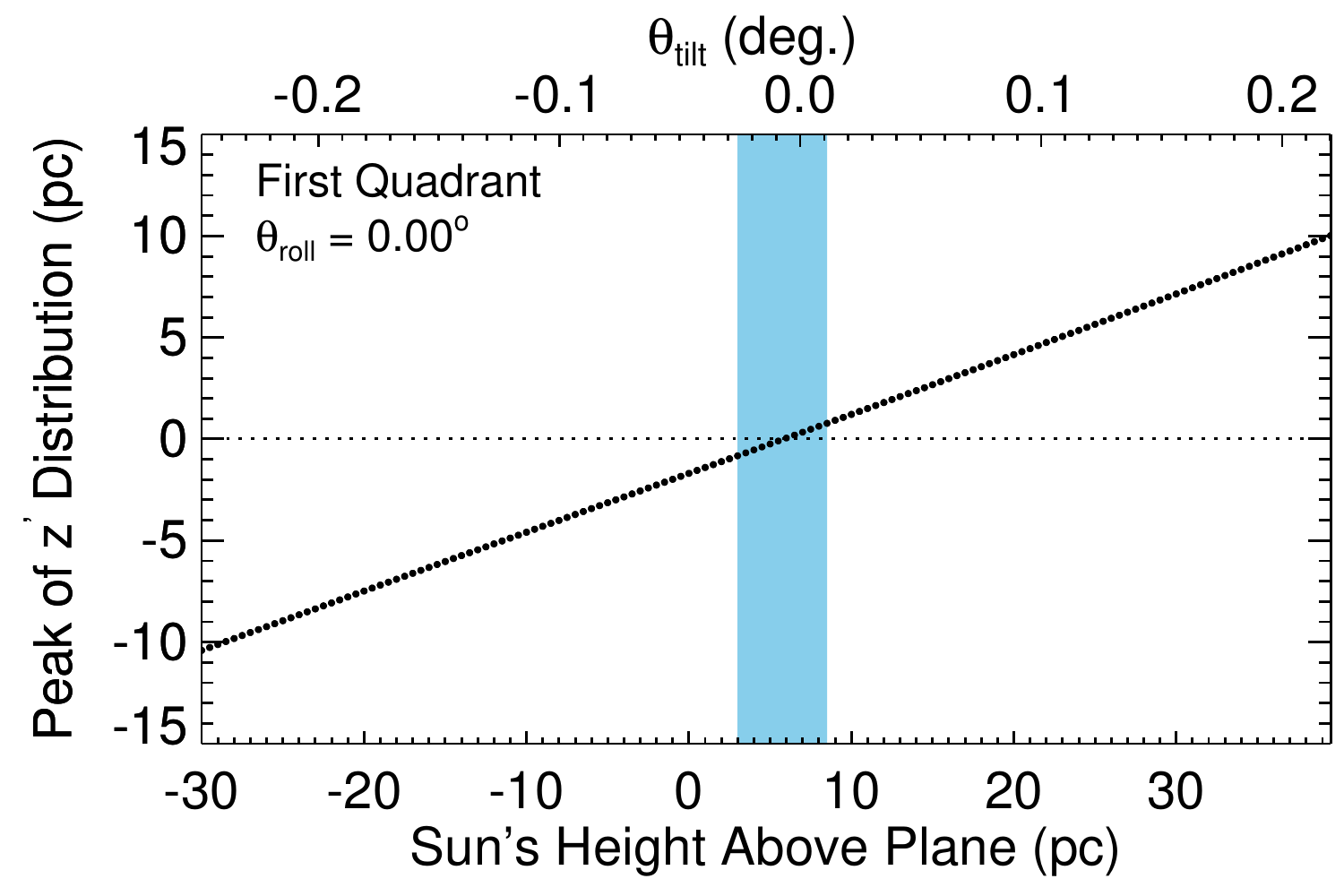}
  \includegraphics[width=3in]{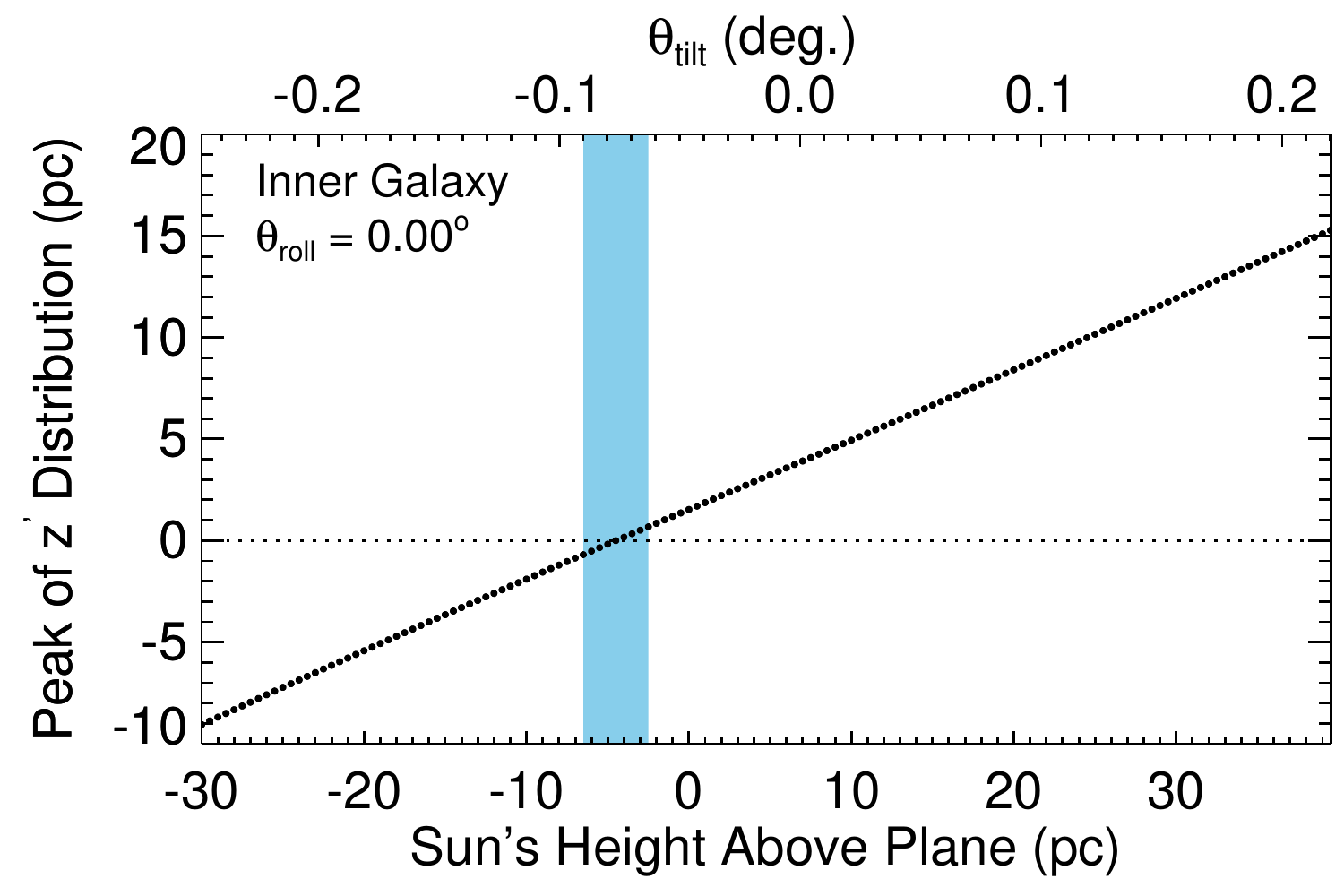}
  \caption{Peak of the \hii\ region distribution as a function of
    solar height above the plane, $z^{\prime}_\sun$ for the first quadrant
    (left) and inner Galaxy (right) samples.  The intersection of the
    blue shaded region with the x-axis shows values of
    $z^\prime_\odot$ that result in \hii\ region distributions that
    peak from $-1$ to $1\,\pc$.
    \label{fig:zprime}}
\end{figure*}

We vary $z^\prime_\sun$ from
$-30\,\pc$ to $40\,\pc$ in steps of $0.5\,\pc$ and recompute
$z^\prime$ for all \hii\ regions in the sample.  For each distribution
of $z^\prime$ values (for a given value of $z^\prime_\odot$), we fit a
Gaussian function to the KDE to determine the peak of the distribution
(as in Figure~\ref{fig:zdist}).  We show the values of the fitted
Gaussian peaks as a function of $z^\prime_\odot$ and $\thetat$ in
Figure~\ref{fig:zprime} and give the derived values of $z^\prime_\odot$ and $\thetat$ in Table~\ref{tab:fits}.

Our analysis finds that the Sun lies
$\sim 5.6\pm2.6\,\pc$ above the HMSF midplane for the first quadrant
sample, for tilt angles of $\sim\!-0.01\degree$, and $4.4\pm1.9\,\pc$
below the plane for the inner Galaxy sample, for tilt angles of
$\sim -0.07\degree$.
The uncertainties in the derived values of $z_\odot^\prime$ and \thetat\ come from allowing the peak to fall within the range $\pm1$\,pc.


\subsection{Midplane Tilt and Roll}
Similar to our investigation of the midplane tilt, we can test for the
midplane roll
by fitting the distributions of $z^\prime$ for the \hii\ region sample.
For a one-dimensional fit, we cannot allow \thetat\ (or, equivalently,
$z_\odot^\prime$) and \thetar\ to both be free parameters.  Instead,
we set \thetar\ to a range of discrete values, and fit for \thetat.

\begin{figure*}
  \centering
  \includegraphics[width=3in]{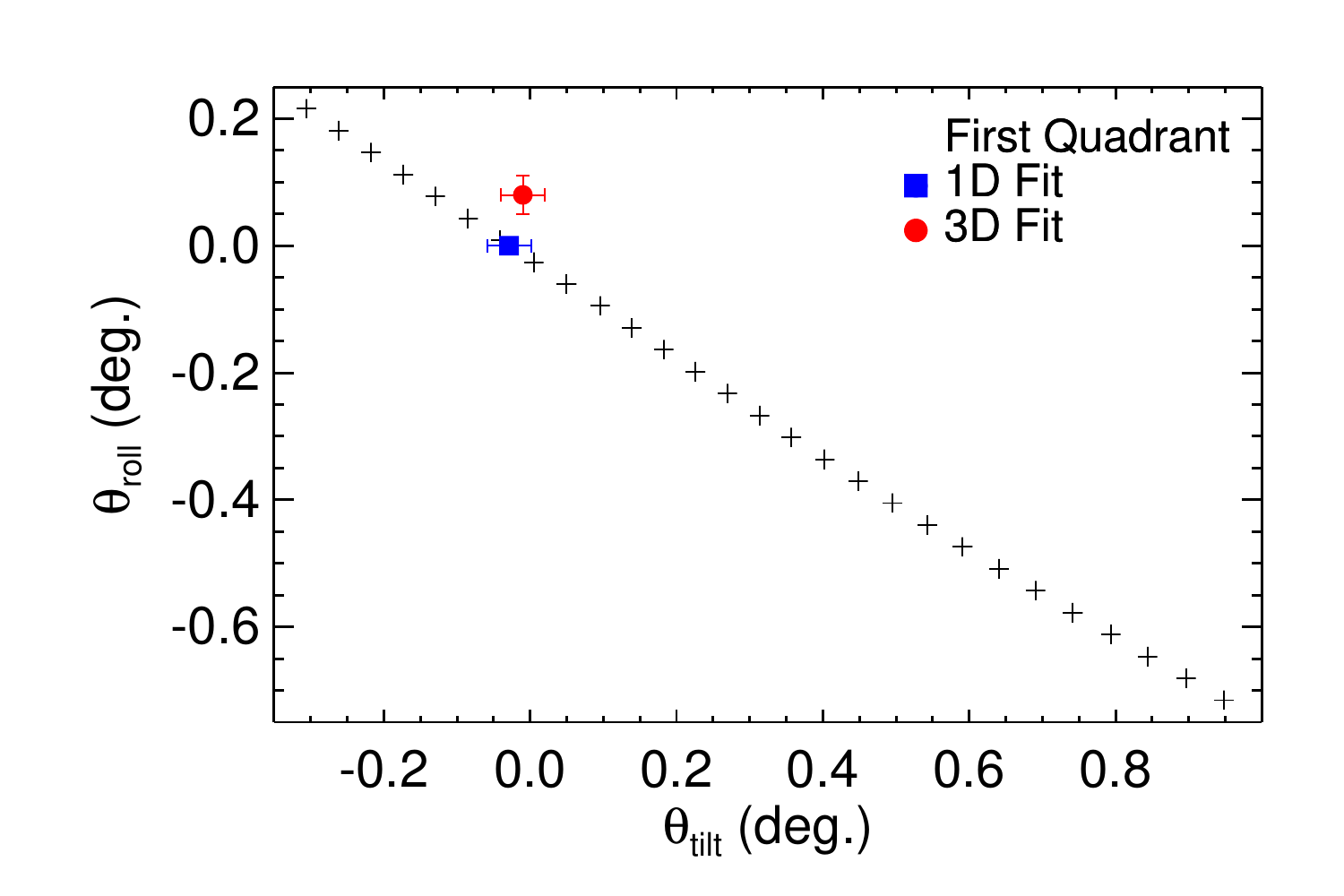}
  \includegraphics[width=3in]{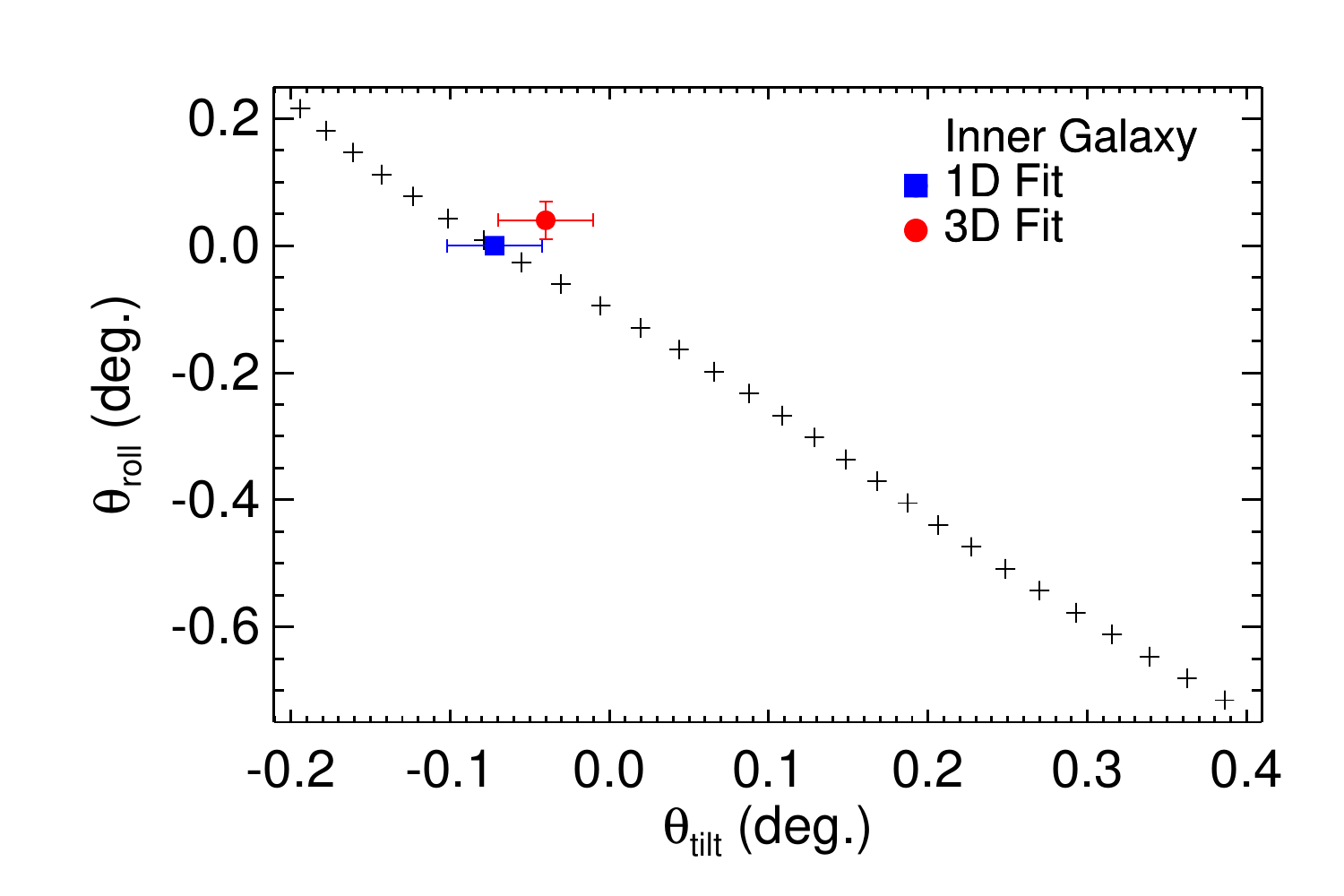}
  \caption{Variations in $\thetar$ as a function of $\thetat$ for the
    first quadrant (left) and inner Galaxy (right) samples.  Each
    point corresponds to the two tilt angles where the
    $z^\prime$-distribution of \hii\ regions is centered at zero.
    There is a negative correlation between the two
    angles.  Blue squares show the values of \thetat\ we derived from one-dimensional fits with $\thetar=0$, whereas the red circles indicate results from three-dimensional fits explained later in Section~\ref{sec:3d}.  Error bars are $3\sigma$.\label{fig:thetas}}
\end{figure*}

We compute a grid of $z^\prime$ distributions for $\thetar$ values
from $-0.8$ to $0.2\degree$ in increments of $0.05\degree$.  We
then vary $z_\odot^\prime$ (and hence \thetat) as in
Section~\ref{sec:tilt} until we find the combination of $\thetat$ and
$\thetar$ where the $z^\prime$ distribution of \hii\ regions peaks at
0\,\pc.  We show the combinations of \thetat\ and \thetar\ that
together lead to distributions peaking at $z^\prime=0$ in
Figure~\ref{fig:thetas} for the first quadrant and inner Galaxy samples.  There is a negative correlation between
\thetat\ and \thetar, such that increasing either angle has the same
effect on the \hii\ region distribution.  From this one-dimensional
analysis we cannot determine the best combination of the two angles,
but we can constrain one angle given the other one.

\subsection{Three-Dimensional Fit for Midplane Tilt and Roll\label{sec:3d}}
Instead of testing discrete values of $\thetat$ and $\thetar$ by
fitting the one-dimensional distribution of $z^\prime$ for
\hii\ regions, we can also simply fit a plane of the form in Equation~\ref{eq:plane} to the \hii\ region
distribution.  In this way, we simultaneously fit for the Sun's height
and the two midplane rotation angles.  The downsides to this method are
that it is difficult to compare with the one-dimensional fits
frequently done by previous authors, and may produce results inconsistent with those of the one-dimensional fits.

We fit the plane using a robust least-squares minimization routine.  The robust least squares fit reduces the impact of outliers on the fit results by minimizing the ``loss function,'' $\rho(z)$, where $z$ is the squared residuals.  We use a ``soft l1'' loss function, $\rho(z) = \sqrt{1+z^2} - 1,$, which is similar to the ``Huber'' loss function.  As before, we force the plane to pass through the location of \sgra.  
We perform these three-dimensional fits for the same subsamples as the one-dimensional
fits and store the results in the
final three columns of Table~\ref{tab:fits}.

The results from the three-dimensional fits are similar to those found previously in the one-dimensional fits, despite allowing for a second midplane tilt angle.  We find small absolute tilt angles and small positive values for the Solar height about the plane.  In the first quadrant sample, for example, $\thetat = -0.01\degree$ and $z_\sun^\prime = 5.7$.  The roll angle is generally small as well, and almost always positive; $\thetar = 0.08\degree$ for the first quadrant sample and $\thetar - 0.04\degree$ for the inner Galaxy sample.  
As can be seen in Figure~\ref{fig:thetas}, the values of $\thetat$ and $\thetar$ derived from the three-dimensional fits are broadly consistent ($\sim 5\sigma$) with the one-dimensional relationships.
This gives us additional
confidence in the derived values.

\subsection{Effects of Completeness and Latitude Restrictions}
Since the {\it WISE} catalog contains a greater quantity of extremely
distant sources compared with other catalogs of star formation
regions, we here examine if our results would change if the sample
were less complete.  

\subsubsection{Artificial Distance Cutoff}
\begin{figure}
\includegraphics[width=3.3in]{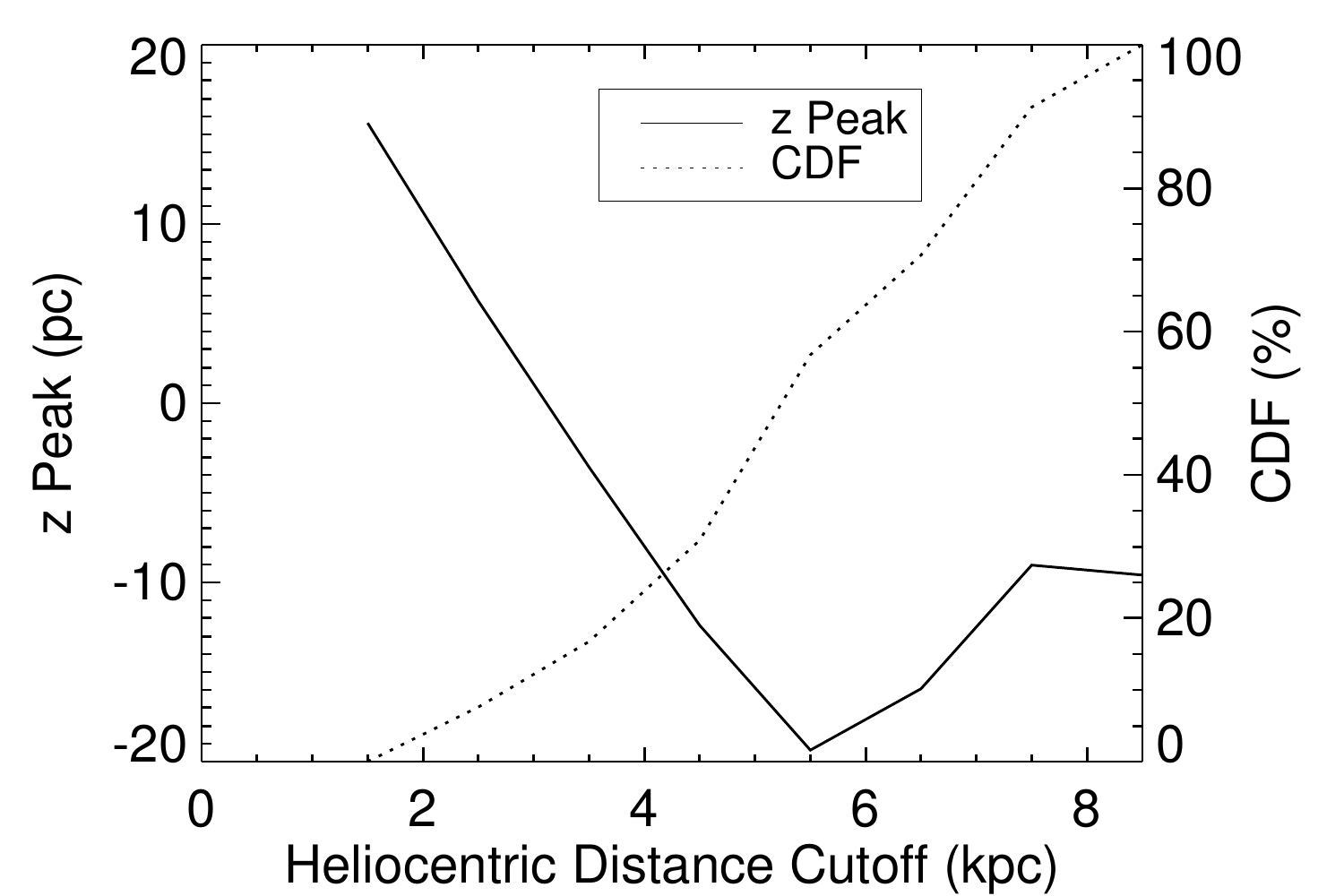}
\caption{Variations in the peak of \hii\ region distributions restricted by Heliocentric distance cutoffs from 1.75 to 8.25\,\pc\ (solid curve).  Also shown is the cumulative distribution
  function for the \hii\ region sample unrestricted by distance (dotted curve). 
  Imposing a distance cutoff
  changes the derived $z$-distribution peak from $-20$ to $+15$\,\pc.
\label{fig:avg_z}}
\end{figure}

As a first test, we restrict our first quadrant
sample by removing sources above a range of Heliocentric distances from 1.75 to 8.25\,\kpc.
We then fit the KDE distributions of these restricted samples with
Gaussians as before, and show the results in Figure~\ref{fig:avg_z}.
We see in this figure that the peak of the $z$ distributions varies from $-19\,\pc$ to $+17\,\pc$.  This simple analysis shows that an artificial distance cutoff may
have a significant effect on the derived values.

\subsubsection{Malmquist Bias}
A flux-limited distribution of course does not have a hard distance cutoff.  We attempt to model a more accurate Malmquist bias by comparing the \hii\ region distance distribution to that of another flux-limited sample.  We choose to do the comparison with the Bolocam
Galactic Plane Survey (BGPS) catalog \citep{ellsworth-bowers15}, although any flux-limited sample could serve the same purpose. 

The BGPS
catalog contains 3508\,mm clumps, 1710 of which have kinematic
distances.  
In the first quadrant zone $10\degree <
\ell < 75\degree$, there are 2843 clumps identified from mm-wave observations, 1214 of which have
kinematic distances.  
The distribution of BGPS
Heliocentric distances differs from that of the {\it WISE} \hii\ regions in
that it has a stronger peak near 5\,\kpc\ and a steeper decrease
thereafter (see below).  This indicates that there is either an
asymmetry in the mm-clump/\hii\ region ratio or that there is
Malmquist bias in the subset of BGPS sources with known distances.


We attempt to evaluate the impact of potential Malmquist bias by
applying a source removal function to the \hii\ region sample
that more closely approximates the effects of Malmquist bias.  
We create a modified \hii\ region distribution
by keeping all sources in the catalog that have Heliocentric distances less
than a cutoff distance $d_{\rm cut}$.  For sources with distances
greater than the cutoff value, we apply a power law source removal
function with a power law index $\alpha$.  The percent likelihood that
a source with a Heliocentric distance $d$ is kept in the catalog is
therefore:
\begin{equation}
    p =\begin{cases}
       1, &\mbox{if } d < d_{\rm cut} \\
       (d_\odot - d_{\rm cut})^{-\alpha},&\mbox{if } d \ge d_{\rm cut}\,.
    \end{cases}
\end{equation}
We iterate $d_{\rm cut}$ and $\alpha$ in the respective ranges
0 to 12\,\kpc\ and 0 to 1 to create modified \hii\ region distributions,
and perform a Kolmogorov-Smirnov (K-S) test on the BGPS and modified
\hii\ region distributions.  The K-S test can determine the likelihood
that two samples are drawn from the same parent distribution.

\begin{figure}
\includegraphics[width=3.in]{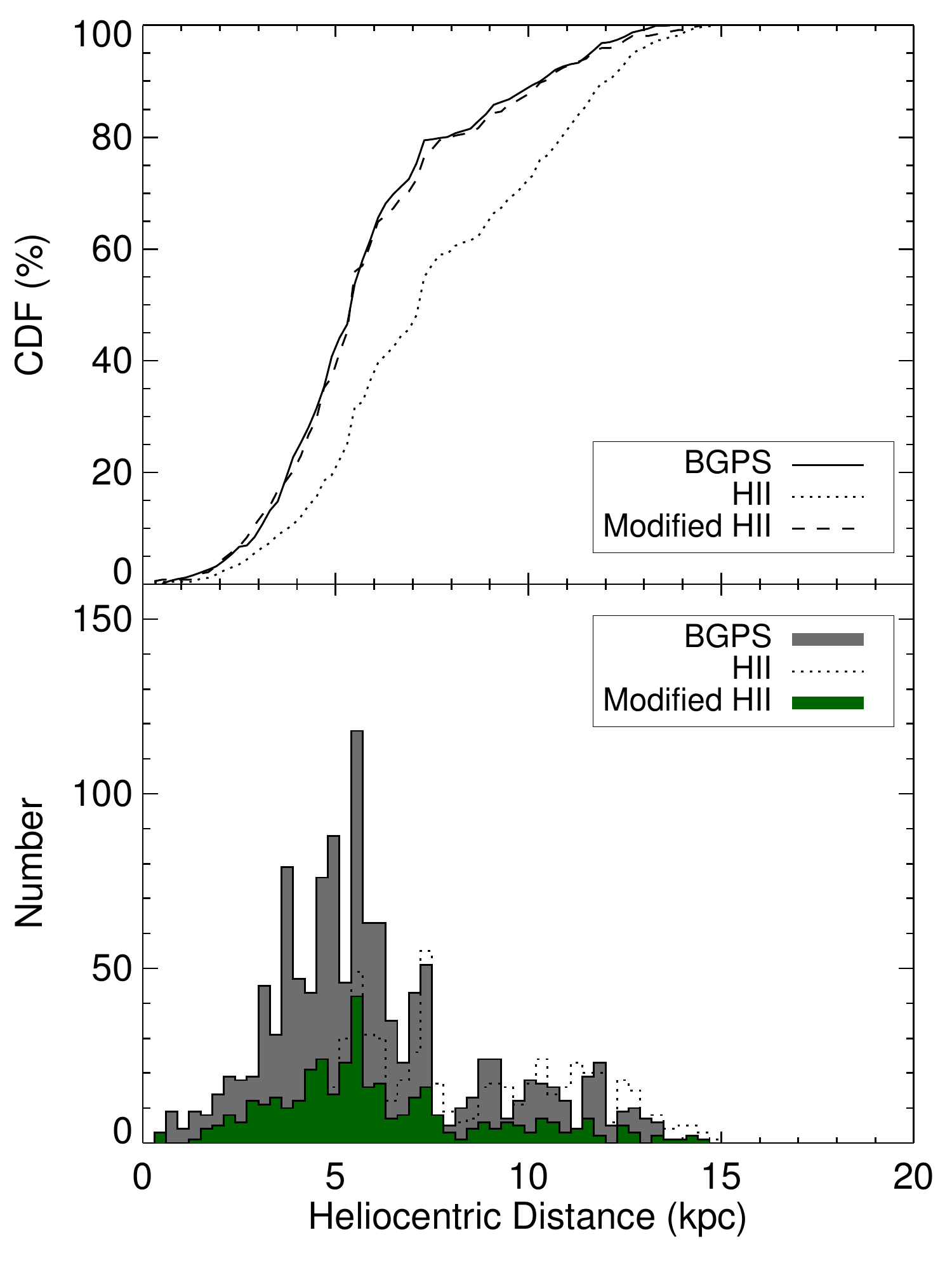}
\caption{First-quadrant distributions of Heliocentric distances
  for the BGPS and \hii\ region samples.  Since the BGPS catalog is
  limited to within $\absb < 0.5\degree$ in the first quadrant, we
  also show the \hii\ region distribution with this limitation.  The
  BGPS distribution is heavily weighted toward relatively nearby
  sources compared to the \hii\ region distribution.
\label{fig:bgps_modified}}
\end{figure}

We find that the two distributions are most similar when $\alpha =
-0.66$ and $d_{\rm cut} = 4.7\,\kpc$.  
We show the Heliocentric distance
cumulative distribution functions (CDFs) before and after the modification in
Figure~\ref{fig:bgps_modified}, top panel, and the distributions
themselves in Figure~\ref{fig:bgps_modified}, bottom panel.

Fitting this modified \hii\ region distribution using one-dimensional fits as in Section~\ref{sec:wise} does impact the derived parameters, although the fits to the modified distribution give similar peaks and scale heights as are found for the complete distribution.  Repeating the Sun's height
analysis, we find that the modified \hii\ region distribution is
consistent with the Sun lying 12.5\,\pc\ above the midplane, for a tilt angle $\thetat = 0.04\degree$.  These values are considerably larger than the unmodified
values of 5.5\,\pc\ and $\thetat = -0.01\degree$.  We conclude that Malmquist
bias can significantly alter the derived values of the Galactic
midplane.

\subsubsection{Latitude Restrictions\label{sec:latitude}}
Most surveys of the Galactic plane are restricted in latitude.  For
example, the BGPS was limited to $\absb < 0.5\degree$.  When creating
the {\it WISE} catalog, we searched {\it WISE} data within $8\degree$ of the
plane, but included known regions outside this range.  The {\it WISE}
catalog therefore is more complete in latitude compared with catalogs
derived from most other Galactic plane surveys.  
To test the effect of this latitude limitation, we restrict our
first-quadrant sample to within $\absb < 0.5\degree$ and $\absb < 1.0\degree$ and repeat the
above analyses.  We again find the peak and scale height are essentially unchanged.  The solar height is $3.0\,\pc$ for $\absb < 0.5\degree$ and $6.5\,\pc$ for $\absb < 1.0\degree$, for tilt angles $\thetat = -0.03\degree$ and $\thetat = 0.00\degree$, respectively.  The former values are considerably different from the unrestricted
values of 5.8\,\pc\ and $\thetat = -0.01\degree$.  Therefore, restricting the sample to within $\absb < 0.5\degree$
significantly changes our results, but there is no such effect if limited
to $\absb < 1.0\degree$.  We conclude that surveys with $\absb < 1.0\degree$ can reproduce the $z$-distribution results from our more complete sample, but surveys with $\absb < 0.5\degree$ cannot.

\section{Summary}
We developed a framework for studies of the Galactic midplane, assuming that the midplane passes through the location of \sgra.  We allowed for rotation of the midplane about Galactic azimuths of $90\degree$ (the ``tilt''; \thetat) and $0\degree$ (the ``roll''; \thetar).  Our framework can be applied to any sample of Galactic objects to determine the midplane, thereby determining the tilt and roll angles with respect to the current midplane definition and also the Sun's height above the midplane.

We applied this framework to the {\it WISE} Catalog of Galactic \hii\ Regions to define the high-mass star formation (HMSF)
midplane.  
In other work (Armentrout et al., 2018, in prep.), we have found that
the {\it WISE} catalog is statistically complete for all
first-quadrant \hii\ regions ionized by single O-stars, giving us a
volume-limited sample.  The fourth quadrant sample is less complete,
and we therefore analyze the first quadrant and inner Galaxy (first
and fourth quadrants) samples independently.  
We computed a Galactic latitude \hii\ region scale height of $\sim 0.30\degree$, and a $z$-distribution scale height of $\sim 30\,\pc$.  These values are dependent on the size of
the \hii\ regions themselves.  The smallest \hii\ regions ($<2\,\pc$
radius) have the smallest scale height distribution of $26\,\pc$.
Larger \hii\ regions have a scale height of $\sim 40\,\pc$.  Since the
size of an \hii\ region depends on its age, these results may indicate
a broadening of the \hii\ region distribution as the regions
themselves evolve.

We found that the HMSF midplane is not significantly different from the current IAU midplane, and that the Sun is near to the HMSF midplane.  Values for the first quadrant and inner Galaxy samples are similar, although the first quadrant sample analysis supports a solar height of $\sim\,5$\,pc above the current midplane and the inner Galaxy sample analysis supports a solar height of a few pc below the current midplane.
The tilt
and roll angles as defined here are negatively correlated, although when $\thetar$ is a free parameter we find similar values for $\thetat$ and the solar height as when $\thetar$ is set to zero. We caution that the roll angle is not well-constrained due to a lack of \hii\ regions with known distances in the fourth Galactic quadrant.  

Our values for the solar height are $\sim 15$\,\pc\ less than
those found in studies of stars, but they are consistent with many results of HMSF tracers.
The meaning of this discrepancy is unclear.  The stellar samples are compiled over a different portion of the Galaxy compared to ours, since extinction drastically limits the distance stars can be seen in the midplane.  The discrepancy between our results may indicate that near to the Sun there is a large-scale displacement in the stellar population.  Since many of the stellar studies rely on counting stars toward the north and south Galactic poles, asymmetries in the stellar distribution would alter the derived result for the solar height.  For example, \citet{xu15} discovered an asymmetry in the main-sequence star counts using data from the Sloan Digital Sky Survey such that there are more stars in the north 2\,\kpc\ from the Sun, more stars in the south between $4-6\,\kpc$ from the Sun, and more stars in the north between $8-10\,\kpc$ from the Sun.  Such asymmetries may make determinations of the solar height difficult using star counts.

Our values are, however, broadly consistent with the
results of \citet{wegg15}.  They found from near-infrared star counts
of red clump giants that the mean latitude in the Galactic long bar is
$b \simeq -0.1\degree$ and that the long bar lies in the midplane after accounting for the midplane tilt.  
At a longitude of the end of the long bar of $\ell \simeq 30\degree$, a distance of 6000\,\kpc, $b = -0.10\degree$, $\thetat = -0.01\degree$, and $\thetar = 0\degree$, $z^\prime \simeq -4\,\pc$.  If we redo the calculation with the first-quadrant sample values of $\thetat = -0.01\degree$ and $\thetar = 0.08\degree$, $z^\prime \simeq -0.1\,\pc$, which implies that the long bar is exactly in the modified midplane.  Using the inner Galaxy values of $\thetat = -0.04\degree$ and $\thetar = 0.04\degree$, $z^\prime \simeq -4\,\pc$.

We tested the robustness of these results and applicability of our
methodology using various permutations of our sample.  Since most
Galactic plane surveys are restricted in latitude, we examine
\hii\ region samples restricted to within $\absb < 1.0\degree$ and
$\absb < 0.5\degree$.  We found that the $\absb <
0.5\degree$ sample results are considerably different, indicating that similar studies of the Milky Way vertical distribution should ideally not include a latitude restriction, and certainly cannot be limited to $\absb < 0.5\degree$. Introducing an
artificial Malmquist bias also changes the results significantly.

\begin{acknowledgments}
  This work is supported by NSF grant AST1516021 to LDA.  TMB acknowledges support from NSF grant AST 1714688.  Support for
  TVW was provided by the NSF through the Grote Reber Fellowship
  Program administered by Associated Universities, Inc./National Radio
  Astronomy Observatory.  We thank Bob Benjamin for enlightening
  discussions on early drafts of this manuscript.  We thank Virginia Cunningham for help on early analyses for this project.
\nraogboblurb\
\end{acknowledgments} 

\clearpage
\bibliographystyle{aasjournal} 
\bibliography{ref.bib} 

\begin{appendix}
\section{Coordinate Conversions\label{sec:appendix}}
\citet{ellsworth-bowers13} derived the relationship between Galactic
positions measured from the current coordinate system that uses the
IAU standards and the values measured from the modified Galactic
plane.  Their derived relationship did not consider the offset of \sgra from
$\lb = (0\degree, 0\degree)$ or the roll of the plane, and we update
their calculations with these changes.

In our nomenclature, unprimed coordinates refer to the values in the
currently-defined Sun-centered coordinate system and primed values
refer to the \sgra-centered values that include the tilt and roll of
the midplane.
See Figures~\ref{fig:faceon}, \ref{fig:geometry}, and
\ref{fig:geometry_roll} for images of the various distances and
angles.

For local Cartesian coordinates with the Sun at the origin, the
current IAU definition gives:
\begin{equation}
\left( \begin{array}{c} x \\ y \\ z  \end{array} \right) =
\left( \begin{array}{c} d_{_\sun}\ \cos \ell\ \cos b \\ d_{_\sun}\ \sin \ell\ \cos b \\ d_{_\sun}\ \sin b \end{array} \right)\,,
\label{eq:local_coord_app}
\end{equation}
where $\hat{x}$ points from the Sun to the Galactic Center, $\hat{y}$
is in the Galactic plane and points toward a Galactic azimuth of
$90\degree$, and $\hat{z}$ points toward the north Galactic pole.

The location of Sgr~A$^*$ is $(\ell_{\sgramath}, b_{\sgramath}) =
(359.944249, -0.046165\degree)$ \citep{reid04}, so
\begin{equation}
  y_{\rm SgrA^*} = R_0\sin \ell_{\sgramath}\,.
\end{equation}
and
\begin{equation}
  z_{\rm SgrA^*} = R_0\sin b_{\sgramath}\,.
\end{equation}
The tilt angle between the currently-defined Galactic midplane and the
modified plane is:
\begin{equation}
  \thetat = \sin^{-1}\left(\frac{z^\prime_\odot+z_{\sgramath}}{R_0}\right)\,.
\end{equation}
This differs from the angle used in \citet{ellsworth-bowers13} by the
additional term $z_{\sgramath}$.  Note that $z_{\sgramath}$ is negative for $b_{\sgramath} < 0$.

The full translational and rotational matrices resulting from
(1) rotating about the $\hat{z}$-axis so that $\hat{x}$ points toward
the Sun,
(2) moving the origin to be centered on \sgra,
(3) rotating about the $\hat{z}$-axis by the angle
$360\degree-\ell_{\sgramath}$ (or, equivalently, clockwise by
$\ell_{\sgramath}$) so that $\hat{x}^\prime$ points toward the Sun,
(4) rotating the Galactic plane about the ${\hat y}$-axis (with the
Sun at the origin) clockwise by the angle $\thetat$, and
(5) rotating the Galactic plane counterclockwise about the ${\hat
  x}$-axis by the angle $\thetar$:
\begin{eqnarray*}
\left( \begin{array}{c} x^\prime \\ y^\prime \\ z^\prime \\ 1  \end{array} \right) =
\left( \begin{array}{cccc} 1 & 0 & 0 & 0 \\
                           0 & \cos\thetar & \sin\thetar & 0 \\
                           0 & -\sin \thetar & \cos \thetar & 0 \\
                           0 & 0 & 0 & 1 \end{array} \right) 
\left( \begin{array}{cccc} \cos\thetat & 0 & -\sin\thetat & 0 \\
                           0 & 1 & 0 & 0 \\
                           \sin \thetat & 0 & \cos \thetat & 0 \\
                           0 & 0 & 0 & 1 \end{array} \right)\\
\left( \begin{array}{cccc} \cos\ell_{\sgramath} & \sin\ell_{\sgramath} & 0 & 0 \\
                           -\sin\ell_{\sgramath} & \cos\ell_{\sgramath} & 0 & 0 \\
                           0 & 0 & 1 & 0 \\
                           0 & 0 & 0 & 1 \end{array} \right)
\left( \begin{array}{cccc} 1 & 0 & 0 & R_0 \\
                           0 & 1 & 0 & y_{\sgramath} \\
                           0 & 0 & 1 & -z_{\sgramath} \\
                           0 & 0 & 0 & 1 \end{array} \right)
\left( \begin{array}{cccc} -1 & 0 & 0 & 0 \\
                           0 & -1 & 0 & 0 \\
                           0 & 0 & 1 & 0 \\
                           0 & 0 & 0 & 1 \end{array} \right)
\left( \begin{array}{c} x \\ y \\ z \\ 1 \end{array} \right)\,.
\end{eqnarray*}
Although the tilt rotation is defined in reference to the current
midplane definition (a rotation centered on the Sun), the roll is
defined from the tilted modified midplane (a rotation centered on
\sgra).  The derived coordinates are therefore:
%
\begin{footnotesize}
\begin{equation}
\left( \begin{array}{c} x^\prime \\ y^\prime \\\\ z^\prime\\\\  \end{array} \right) =
\left( \begin{array}{c} 
  (R_0-x)\cos\ell_{\sgramath}\cos\thetat + (y_{\sgramath}-y)\sin\ell_{\sgramath} \cos\thetat - (z-z_{\sgramath})\sin\thetat\\
  (R_0-x)(\sin\ell_{\sgramath}\cos\thetar + \cos\ell_{\sgramath} \sin\thetat\sin\thetar) - (y_{\sgramath}-y)(\cos\ell_{\sgramath} \cos\thetar -\\ \sin\theta\ell_{\sgramath} \sin_{\rm tilt} \sin_{\rm roll}) + (z-z_{\sgramath})\cos\thetat\sin\thetar\\
  (R_0-x)(\sin\ell_{\sgramath}\sin\thetar - \cos\ell_{\sgramath} \sin\thetat\cos\thetar) + (y_{\sgramath}-y)(\sin\ell_{\sgramath} \cos\thetar \sin\thetat + \cos\ell_{\sgramath} \sin\thetar) +\\ (z-z_{\sgramath})\cos\thetat\cos\thetar\\
\end{array} \right) \, .
\label{eq:xyz_gal}
\end{equation}
\end{footnotesize}
Since $\sin(\ell_{\sgramath}) = -0.0009730 \simeq 0$ and $\cos(\ell_{\sgramath}) = 0.9999995 \simeq 1$, these reduce to
\begin{footnotesize}
\begin{equation}
\left( \begin{array}{c} x^\prime \\ y^\prime \\ z^\prime  \end{array} \right) =
\left( \begin{array}{c} 
  (R_0-x)\cos\thetat  - (z-z_{\sgramath})\sin\thetat\\
  (R_0-x)\sin\thetat\sin\thetar + (y_{\sgramath}-y)\cos\thetar + (z-z_{\sgramath})\cos\thetat\sin\thetar\\
  (R_0-x)\sin\thetat\cos\thetar - (y_{\sgramath}-y)\sin\thetar + (z-z_{\sgramath})\cos\thetat\cos\thetar\\
\end{array} \right) \, .
\label{eq:xyz_gal2}
\end{equation}
\end{footnotesize}
Since $x$ and $y$ are of order $\sim\!\kpc$ and because $\theta_{\rm
  tilt}$ and $\thetar$ are both small, rotations of the midplane will
have a larger fractional affect on derived values in $z^\prime$
compared to $x^\prime$ and $y^\prime$.  In the limit that $\thetar =
0$,
\begin{footnotesize}
\begin{equation}
\left( \begin{array}{c} x^\prime \\ y^\prime \\ z^\prime  \end{array} \right) =
\left( \begin{array}{c} 
  (R_0-x)\cos\thetat - (z-z_{\sgramath})\sin\thetat\\
  y_{\sgramath}-y\\
  (R_0-x)\sin\thetat + (z-z_{\sgramath})\cos\thetat\\
\end{array} \right) \, .
\label{eq:xyz_gal_simp}
\end{equation}
\end{footnotesize}
These expressions differ from those in \citet{ellsworth-bowers13} by
the additional $y_{\rm \sgramath}$ and $z_{\rm \sgramath}$ terms.

\end{appendix}



\end{document}